\documentclass[a4paper,fleqn]{cas-dc}

\usepackage[numbers]{natbib}

\def\tsc#1{\csdef{#1}{\textsc{\lowercase{#1}}\xspace}}
\tsc{WGM}
\tsc{QE}
\tsc{EP}
\tsc{PMS}
\tsc{BEC}
\tsc{DE}

\usepackage[normalem]{ulem}
\newcommand{\stkout}[1]{\ifmmode\text{\sout{\ensuremath{#1}}}\else\sout{#1}\fi}

\begin{document}
\let\WriteBookmarks\relax
\def\floatpagepagefraction{1}
\def\textpagefraction{.001}

\shorttitle{The 3D Trajectory-based Stress Visualizer}

\shortauthors{J Wang et~al.}

\title[mode = title]{3D-TSV: The 3D Trajectory-based Stress Visualizer}                      



%
\author[1]{Junpeng Wang}[
                        orcid=0000-0002-4607-844X]

\cormark[1]


\ead{junpeng.wang@tum.de}


\credit{Conceptualization of this study, Methodology, Writing - Review \& Editing, Software}

\affiliation[1]{organization={Technical University of Munich},
    addressline={Boltzmannstr. 3}, 
    city={Garching},
    postcode={85748}, 
    country={Germany}}

\author[1]{Christoph Neuhauser}[
                        orcid=0000-0002-0290-1991] 
                        
\ead{christoph.neuhauser@tum.de}

\credit{Methodology, Writing - Review \& Editing, Software}

\author[2]{Jun Wu}[%
   orcid=0000-0003-4237-1806]
\cormark[1]
\ead{j.wu-1@tudelft.nl}

\credit{Conceptualization of this study, Methodology}

\affiliation[2]{organization={Delft University of Technology},
    addressline={Landbergstraat 15}, 
    city={Delft},
    postcode={2628 CE}, 
    country={The Netherlands}}

\author[3]{Xifeng Gao}[
                        orcid=0000-0003-0829-7075] 
\ead{xifgao@tencent.com}
\credit{Conceptualization of this study, Methodology}

\affiliation[3]{organization={Lightspeed \& Quantum Game Studios, Tencent America},
    city={Seattle},
    country={USA}}
    
\author[1]{R\"udiger Westermann}[
                        orcid=0000-0002-3394-0731] 
\ead{westermann@tum.de}
\credit{Conceptualization of this study, Methodology, Writing - Review \& Editing, Supervision, Funding acquisition}

\cortext[cor1]{Corresponding author}



\begin{abstract}
We present the 3D Trajectory-based Stress Visualizer (3D-TSV), a visual analysis tool for the exploration of the principal stress directions in 3D solids under load. 3D-TSV provides a modular and generic implementation of key algorithms required for a trajectory-based visual analysis of principal stress directions, including the automatic seeding of space-filling stress lines, their extraction using numerical schemes, their mapping to an effective renderable representation, and rendering options to convey structures with special mechanical properties. In the design of 3D-TSV, several perceptual challenges have been addressed when simultaneously visualizing three mutually orthogonal stress directions via lines. We present a novel algorithm for generating a space-filling and evenly spaced set of mutually orthogonal lines. The algorithm further considers the locations of lines to obtain a more regular appearance, and enables the extraction of a level-of-detail representation with adjustable sparseness of the trajectories along a certain stress direction. To convey ambiguities in the orientation of the principal stress directions, the user can select a combined visualization of two principal directions via oriented ribbons. Additional depth cues improve the perception of the spatial relationships between trajectories. 3D-TSV is accessible to end users via a C++- and OpenGL-based rendering frontend that is seamlessly connected to a MatLab-based extraction backend. 
The code (BSD license) of 3D-TSV as well as scripts to make ANSYS and ABAQUS simulation results accessible to the 3D-TSV backend are publicly available.


\end{abstract}





\maketitle

\section{Introduction} \label{Sec:Intro}
Techniques for visualizing the three mutually orthogonal principal stress directions in 3D solids under load are important in a number of use cases in computational mechanics. In civil engineering such visualizations are used to develop and assess strategies for steel reinforcement of concrete support structures~\cite{tam2015stress}. In mechanical engineering, where often massive components like engines and pumps are considered, one is interested in how forces ``find'' their way through these components. The development of lightweight load bearing structures is investigated in e.g., aerospace engineering, here stress directions provide the first indicators where structures can be hollowed~\cite{kratz2014tensor, kwok2016structural, daynes2017optimisation}. In bio-mechanics, such techniques are used to show tension and compression pathways simultaneously, and compare different structural designs regarding their mechanical properties~\cite{Dick2009Stress}. For an overview of stress tensor visualization, we refer to the recent review article by Hergl et al.~\cite{hergl2021visualization}.

An informative visualization of the stress directions in a 3D solid can be achieved via principal stress lines (PSLs), i.e., integral curves in 3D space along the principal stress directions. PSLs are effective in communicating the pathways along which external loads are transmitted, and they show the mutual relationships between the different principal stress directions~\cite{Dick2009Stress,Wang2020Globally}. In computational engineering, PSLs are used in particular to show where and how loads are internally redirected and deflected. Such visualizations are necessary for a first qualitative analysis, before a quantitative analysis of certain regions using derived scalar stress measures is commonly performed. 

However, in computational mechanics stress trajectory visualizations are used in a rather inconsistent way, and, to the best of our knowledge, no standard tool for such an analysis exists. In many research groups in computational mechanics, own software packages for showing one particular principal stress direction starting at randomly selected locations are used. Often, CFD tools for flow visualization are used to show streamlines in a single principal stress direction field. Visualization tools that are able to show all principal stress directions simultaneously are rare, and also available post-processing tools do not offer this functionality.  

One reason preventing a wider adoption of such tools is visual clutter and occlusions that are produced when showing the different types of PSLs simultaneously. Due to their mutual orthogonality, the visualizations appear irregular and unstructured, and perceptual coherence breaks up even for sparse sets of trajectories. While this effect can be reduced by starting trajectories from narrow regions and following only a single type of PSLs, this leaves large sub-domains uncovered and does not show the mutual variations of the stress directions. In general, clutter can be reduced by visualizing the single stress directions side-by-side, yet juxtaposition makes it difficult to effectively relate the three mutual orthogonal stress directions to each other.
\bigskip
\linebreak
\noindent
\textbf{Contribution}

\begin{figure*}[ht]
  \centering
  \includegraphics[width=0.98\linewidth,trim=0.05cm 6.5cm 1.3cm 0.05cm, clip=true]{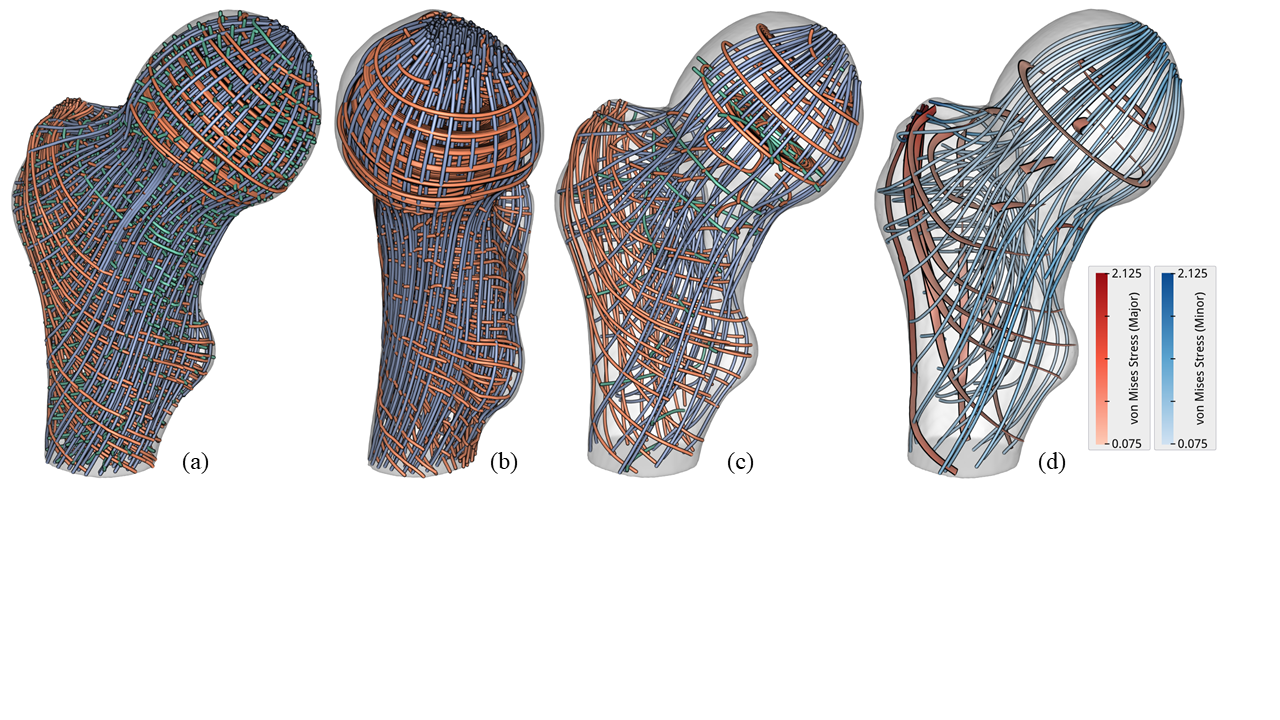} 
  \caption{(a) The 3D Trajectory-based Stress Visualizer generates a space-filling and evenly spaced set of principal stress lines (PSLs) in a 3D domain. (b) It supports a regular appearance by considering already selected lines when locating new seed points. (c,d) To reduce clutter, the density of PSLs can be adapted in a hierarchical manner. (d) Ambiguities in the assignment of stress types to directions are visualized by merging two principal stress directions into ribbons. Different scalar stress measures (d) can be mapped to color.
  }
  \label{fig:teaser}
\end{figure*}

\noindent This paper presents the 3D Trajectory-based Stress Visualizer (3D-TSV), a system and methodology for the visual analysis of the PSLs in 3D stress fields. \autoref{fig:teaser} gives an overview of the visualization options provided by 3D-TSV. With 3D-TSV, we release a system that supports a comprehensive integral line-based analysis of 3D stress fields. To achieve this, 3D-TSV builds upon existing techniques for line seeding in vector fields~\cite{Jobard1997creating,IlluminatedStreamlines}, and it extends them towards the specific use case by considering simultaneously the three principal stress directions in the seeding process.
3D-TSV is designed to achieve improved regularity of the extracted PSLs, i.e., it aims for a grid-like structure where PSLs roughly intersect, uniformly cover the domain, and reveal symmetries in the underlying fields. To achieve this, in the sequential seeding process every new seed point is located on an existing PSL belonging to a different principal stress direction. As proposed for streamlines in~\cite{Jobard1997creating,IlluminatedStreamlines}, the seeding process is parameterized using different distance thresholds for each type of PSL, which allows controlling separately the sparseness of the PSLs of each type. We use this possibility to enable a level-of-detail (LoD) visualization that combines a dense seeding of a selected PSL type with a seeding at a user-selected sparseness level of the respective other PSLs. 

To ease integration into existing systems and accessibility to end users, 3D-TSV is implemented as a client-server tool connecting a MatLab PSL extraction backend with an OpenGL rendering frontend. The backend extracts trajectories from a given stress field using parameters that are either specified via the GUI that is built into the renderer, or a configuration file. We have chosen a MatLab backend due to the popularity of MatLab in mechanical engineering, and, thus, to enable engineers to easily integrate new model representations and algorithms. Currently, 3D-TSV works with hexahedral simulation grids, including MatLab code for trilinear and inverse distance-based interpolation of stress tensors in such grids. If other types of basis functions are used, the corresponding MatLab functions simply need to be exchanged. 
Due to the cell adjacency structure that is built internally to efficiently find the next cell during trajectory integration in deformed hexahedral grids, other cell types can be supported with only minor additional effort.

The frontend renders whatever set of lines that is sent from the backend using advanced rendering options such as depth cues, outlines, as well as ambient occlusion effects to improve the perception of the spatial relationships between trajectories. Furthermore, the user can select to visualize one pair of stress directions via ribbons. Ribbons follow one of the selected directions and twist according to the other one, and they can effectively convey regions where the assignment of the eigenvector directions to the type of PSL (i.e., major, medium, or minor) changes.

To summarize, the contributions of this work are
\begin{itemize}
    \item an advanced and publicly available tool for trajectory-based stress tensor visualization supporting stress fields on arbitrary hexahedral grids, 
    \item the adaptation of evenly spaced line seeding to create a space-filling set of PSLs with improved regularity, 
    \item an adaptive level-of-detail visualization using varying PSL density and visual mappings to lines and ribbons.
\end{itemize}


The application of 3D-TSV is demonstrated in a number of experiments using datasets with different shapes and stress states. The code of 3D-TSV is made publicly available under a BSD license, and published on \url{https://github.com/Junpeng-Wang-TUM/3D-TSV}. 
In video1\footnote{\url{https://youtu.be/lN9CxgvfgNY}}, the seeding of trajectories by 3D-TSV is compared to the seeding of trajectories separately in each principal stress direction field via evenly spaced seeding~\cite{Jobard1997creating}. 3D-TSV can be used as client-server system as described (see video2\footnote{\url{https://youtu.be/h7BzP7Jg_-o}}), or as standalone tool solely in MatLab providing rudimentary visualization options (see video3\footnote{\url{https://youtu.be/99Jn938ZoVk}}). Also the frontend can be used standalone, reading the PSL specific information from "psl.dat" files (see video4\footnote{\url{https://youtu.be/zafBOAt9Xvs}}). Thus, any other backend can be used to generate PSLs and let the frontend visualize them. We also provide a script written in the ANSYS built-in language APDL, which automatically converts the result of an ANSYS finite element stress analysis into the format required by the 3D-TSV backend (see video5\footnote{\url{https://youtu.be/Yri_B7m3AWU}}). To support the output from ABAQUS, the mesh information needs to be read from the ABAQUS input file (".inp"), and the stress data can be acquired from the result file (".rpt"). We provide datasets, description and configuration files, as well as scripts for all use cases of 3D-TSV on the publicly available GitHub repository. 

\section{Related work}
\textbf{Stress Tensor Field Visualization.}
Stress tensor field visualization can be classified into trajectory-, glyph- and topology-based methods~\cite{Kratz2013Visualization,hergl2021visualization}. Trajectory-based methods choose the PSLs as visual abstractions of the stress field, focusing on the directional structure of the principal stresses. Delmarcelle and Hesselink~\cite{Delmarcelle1993Visualizing} introduced the concept of hyperstreamlines, a visual mapping of the medium and minor principal stresses onto a tube surface with a single selected major PSL as centerline. 
Dick et al.~\cite{Dick2009Stress} trace the major and minor PSLs from randomly distributed seed points in the loading area of the solid object, and different types of stress state like tension and compression are distinguished by color. In order to identify and visualize regions where stress trajectories are of rotational or hyperbolic behavior, Oster et al.~\cite{Oster2018Core} proposed the concept of tensor core lines in 3D second‐order tensor fields. Hotz et al.~\cite{Hotz2006Tensor} smear out dye along the PSLs using line integral convolution. In this way, a density field is generated that resembles a grid-like structure. This approach provides a global overview of a 2D stress distribution, yet an extension to 3D is problematic due to the generation of a dense volumetric field.

It's worth noting that even though stresses are frequently simulated and analysed in engineering applications, the use of trajectory-based visualizations that consider the whole stress field as a tensor field instead of several scalar fields are not commonplace. In particular, such functionality seems neither provided by any of the well-established software packages for stress simulation, like ABAQUS and ANSYS, nor by dedicated environments for visualizing finite-element simulation results~\cite{lee2008femvrml, weng2011web}. 

Glyph-based methods, on the other hand, depict the stress field by a set of well-designed geometric primitives -- so-called tensor glyphs. Tensor glyphs were originally designed for glyph-based diffusion tensor visualization~\cite{Kindlmann2004Superquadric}, and later adapted to visualize positive definite tensors \cite{Kindlmann2008Quantification}, general symmetric tensors~\cite{Schultz2010Superquadric}, as well as asymmetric tensors\cite{Seltzer2016Glyphs,Gerrits2017Glyphs}. Glyph-based techniques are problematic when used to visualize 3D stress fields, due to their inherent occlusion effects. Specific placement strategies can be used to reduce the number of glyphs and occlusions thereof~\cite{Kindlmann2006Diffusion, Hlawitschka2007Interactive}. 
Tensor glyphs are effective in showing the local stress states, but they cannot effectively communicate the global structure of stress lines. Patel and Laidlaw~\cite{patel2020visualization} proposed to guide the placement of glyphs by principal trajectories in the underlying field, and thus to provide a better understanding of the global relationships in this field. 

Topology-based approaches for stress tensor visualization abstract from the depiction of stress directions and focus on revealing specific topological characteristics of the tensor field. Delmarcelle and Hesselink~\cite{Delmarcelle1994Topology, Hesselink1997Topology} studied the topology of symmetric 2D and 3D tensor fields, and introduced the fundamental concepts of degenerate points and topological skeletons. Zheng and Pang ~\cite{Zheng2004Topological}, and later Roy et al.~\cite{Roy2018Robust}, discussed the robust extraction of these topological features. Zobel and Scheuermann proposed the notion of extremal points to analyze the complete invariant part of the tensor~\cite{Zobel2018Extremal}. Raith et al. presented a general approach for the generation of separating surfaces in the invariant space~\cite{Raith2018Tensor}. Palacios et al. introduced the eigenvalue manifold and visualized the 3D eigenvectors as curve surfaces~\cite{Palacios2015Feature}. Qu et al.~\cite{Qu2020Mode} further generalized the concepts of degenerate curves and neutral surfaces to a unified framework called mode surfaces. 


\textbf{Streamline Seeding.}
Seeding strategies to control the density and placement of trajectories in vector fields are widely used in flow visualization. Turk and Banks~\cite{Turk-Stream} and Jobard and Lefer~\cite{Jobard1997creating} were the first to introduce seeding strategies for generating evenly spaced sets of streamlines in 2D vector field. Numerous extensions and improvements of these concepts have been proposed since then. In particular, Vilanova et al.~\cite{Vilanova:2004:DTI} proposed an extension of the approach by Jobard and Lefer to diffusion tensor fields, which detects the distance between the new streamline and the existing ones during the tracing process. They demonstrate the generation of evenly distributed streamlines, however, the approach suffers from `unfinished' streamlines that are caused by an artificial stopping criterion and only considers a single eigenvector field at a time. For 3D flow visualization, dedicated approaches and frameworks have been developed to reduce the visual clutter and occlusion of densely distributed streamlines in 3D fields~\cite{Ye2005Strategy, chen2007similarity, Yu2011Hierarchical, Kanzler2016Line}. However, these techniques do not fit our goal of visualizing PSLs and their mutual relationships, which requires considering three sets of orthogonal PSLs simultaneously. 

\textbf{Streamline Visualization.}
Illuminated streamlines are often used as a means of visualizing streamlines in a 3D environment. The streamlines are mapped to tubes and then shaded, e.g., using the Blinn-Phong shading model \cite{BlinnPhong}. Early work on illuminated streamlines was done by Zöckler et al.~\cite{IllumStreamlines} and Mattausch et al.~\cite{IlluminatedStreamlines}. Stoll et al.~\cite{StylizedLines} extended this work by introducing stylized line primitives, rendered by a hybrid CPU-GPU renderer. Liu~\cite{liu2019prototype} presented the DOXIV, a prototype framework for high-performance visual analysis of large flow data. Volpe~\cite{StreamribbonsInitial} first introduced the concept of streamribbons for flow field visualization.

\textbf{Hexahedral Meshing.}
An alternative approach to PSL-based stress field visualization is to generate a frame field from the principal stress field first and employ field-aligned hexahedral meshing to produce orthogonal edges that follow PSLs. The edges of such hex-meshes can follow the directions of PSLs excellently in situations where degenerate points are not present and the stress lines show low degrees of convergence and divergence. However, when guided with frame fields corresponding to realistic load situations, yet still much more benign than those demonstrated in this work, it is an unsolved problem to reliably produce an all-hex mesh. Hexahedral-dominant meshing has been resorted as an intermediate solution. For instance, Wu et al.~\cite{Wu2021TVCG} propose a conforming stress-guided lattice structure by combining topology optimization with the field-guided polyhedral meshing algorithm from \cite{Gao2017Robust}. Arora et al.~\cite{Arora2018DEsigning} generate similar structural designs via the guidance of the principal stress field, where they modify the stress field to get a smooth frame field. However, hexahedral-dominant meshes often contain either T-junctions or non-hexahedral elements with non-orthgonal edges, significantly deviating from the PSLs and are, thus, not applicable for stress field analysis either.


\section{Stress Tensor Directions} \label{Sec:PrelimWork}
At each point in a 3D solid under load, the stress state is fully described by the stress vectors for three mutually orthogonal orientations. The second-order stress tensor 
\begin{equation} \label{Eqn:StressTensor}
    T = \begin{bmatrix} \sigma_{xx} & \tau_{xy} &\tau_{xz} \\ \tau_{xy} & \sigma_{yy} & \tau_{yz}\\ \tau_{xz} & \tau_{yz} & \sigma_{zz} \end{bmatrix}
\end{equation}
contains these vectors for the axes of a Cartesian coordinate system. 
$T$ is symmetric since the shear stresses given by the off-diagonal elements in $T$ are equal on mutually orthogonal planes. 
The principal stress directions of the stress tensor indicate the three mutually orthogonal directions along which the shear stresses vanish. These directions are given by the eigenvectors of $T$, with magnitudes given by the corresponding eigenvalues. The signs of the principal stress magnitudes classify the stresses into tension (positive sign) or compression (negative sign). However, since there are three principal stresses acting at each point, the classification is with respect to a specific direction.

In descending order, the three eigenvalues of $T$ represent the major $\sigma_1$, medium $\sigma_2$ and minor $\sigma_3$ principal stresses, with the corresponding eigenvectors indicating the principal stress directions at each point in the 3D solid. The trajectories along these directions are called the principal stress lines (PSLs). They are computed by numerically integrating massless particles in each single (normalized) eigenvector field.  

In general, $\sigma_1$, $\sigma_2$ and $\sigma_3$ are mutually unequal, and the eigenvectors are linearly independent and even mutually orthogonal due to the symmetry of $T$. However, so-called degenerate points can exist where two or more eigenvalues are equal. In the vicinity of these points, which are classified by $\sigma_1 = \sigma_2 > \sigma_3$ or $\sigma_1 > \sigma_2 = \sigma_3$\footnote{We do not consider triple degenerate points with $\sigma_1 = \sigma_2 = \sigma_3$, since they do not exist under structurally stable conditions ~\cite{Zheng2004Topological}.}, the PSL direction cannot be decided. Therefore, when tracing along a principal stress direction, we test whether the eigenvalue $\sigma_i$ corresponding to this direction is too close to another eigenvalue $\sigma_j$, i.e., $deg = \frac{1}{2} \left| \frac{\sigma_i-\sigma_j}{\sigma_i+\sigma_j} \right| < 10^{-6}$. If this is the case and the angle between the PSL tangents at the current and next integration point is too large, the integration is stopped. Furthermore, we provide the option to map $deg$ to the color of a PSL via a color table (see Sec.~\ref{SubSec:PSLgeo}), so that the proximity to a degenerate point is indicated. PSL integration is also stopped when the next integration point is located on a boundary face, the point is closer to a previous point on the same trajectory than a predefined distance threshold (i.e., to avoid running into closed orbits), or the number of integration steps reaches the pre-defined threshold. 

The integration of PSLs requires to select seed points from which they start until they arrive at a degenerate point or the boundary. While uniform seeding in the entire domain is used as the default option, the user can select seeding from the boundary vertices as well as the vertices where loads are applied. Furthermore, different integration schemes can be used for PSL tracing, including the 1st-order Euler method, and the 2nd- and 4th-order Runge-Kutta methods, where the fixed integration step size $\delta$ is used for Cartesian meshes, and an adaptive $\delta$ for unstructured hexahedral meshes. In each integration step, the stress tensor $T$ is interpolated, and the eigenvalues and eigenvectors are computed from the interpolated tensor. If none of the mentioned stopping criteria holds, the next step is performed in the direction with the least deviation from the previous direction. 


\section{PSL Seeding and Level of Detail} \label{Sec:seeding-lod}
Finding a set of PSLs that effectively convey the principal stress directions in 3D stress fields requires to consider perceptual issues related to the visualization of large sets of trajectories. While in principle the PSLs of a single type, i.e., major, medium, or minor, can be visualized separately using techniques from flow visualization, in a stress field the different types of PSLs need to be shown simultaneously to understand their mutual interplay. However, an effective and efficient visual analysis is hindered by the mutual orthogonality of the different types, which is perceived as a disordered state even when a low number of PSLs is shown. Our proposed seeding strategy cannot completely avoid this problem, but it has some built-in regularity due to enforced PSL intersections. 

\begin{figure*}[ht]
    \centering
    \includegraphics[width=0.98\linewidth,trim=0.05cm 4cm 0.05cm 4cm, clip=true]{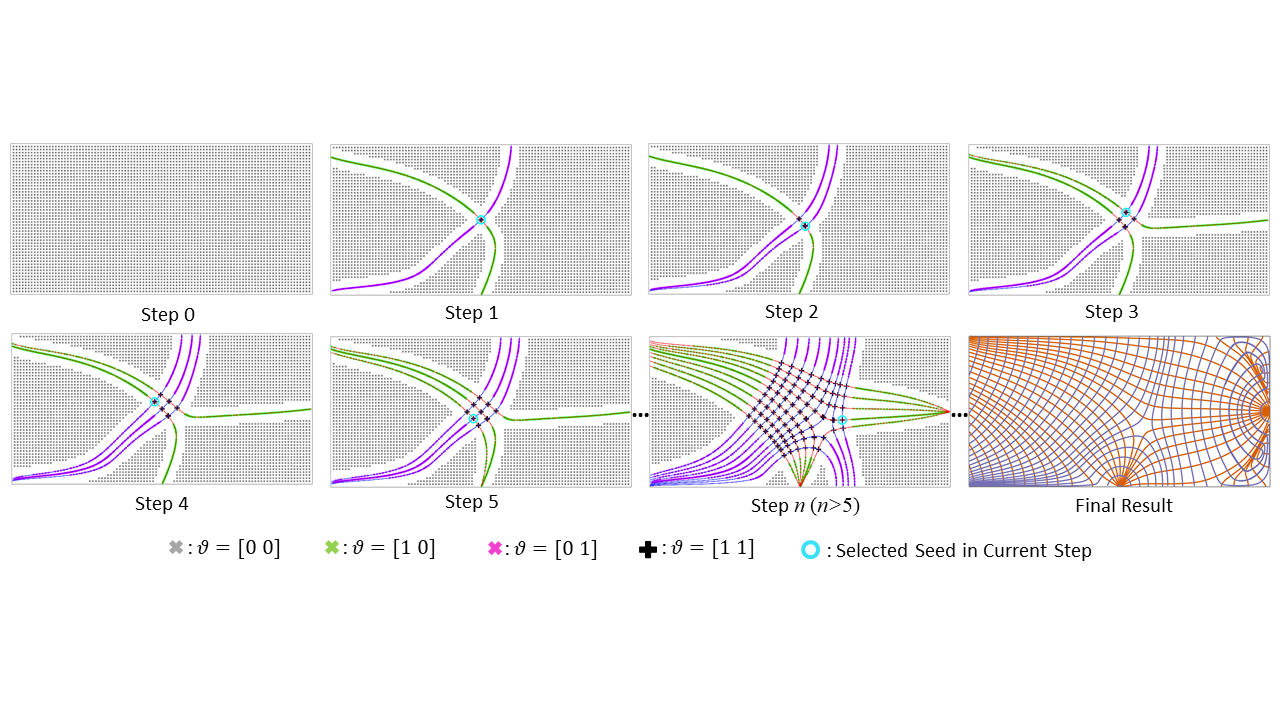}
    \caption{Starting from a set of seeds with empty valence $[0\:\:0]$, the sampling process is performed until all the seed valences have been turned to $[1\:\:1]$. The ocher and blue lines are the major and minor PSLs.}
    \label{fig:methodOverview}
\end{figure*}

\subsection{Evenly Spaced PSL Seeding} \label{SubSec:seeding}
The proposed seeding strategy builds upon the evenly spaced streamline seeding approach by Jobard and Lefer~\cite{Jobard1997creating}, and extends this approach in several ways to account for the application to PSLs. For the sake of clarity, we describe the strategy in the context of 2D stress fields, yet it will become clear that the extension to 3D is straightforward. However, when applied in 3D, the resulting PSL structures show a fundamental difference. Unlike in 2D, where due to the intersections between major and minor PSLs a fairly regular grid-like structure is generated, such intersections are rare or do not exist at all when seeding PSLs in 3D. This counteracts the impression of a consistent grid-like structure and results in a rather disordered appearance. We propose a seeding strategy that weakens this effect, but it needs to be considered that due to the nature of PSLs in 3D stress fields a globally consistent 3D grid-like structure is impossible to achieve in general.  

Our method builds upon the selection of new seed points in the spirit of Jobard and Lefer, where the potential candidates are those points which are at least a prescribed distance away from any already extracted PSL. Of these candidates, the one with minimum distance is selected and a new trajectory is started at that point. In contrast, in our approach the distance is always wrt.~the initial seed point, so that the PSLs grow around that point instead of being seeded at vastly different locations. 

To adapt the seeding strategy to the situation of different types of PSLs, we first introduce the concept of \emph{seed valence}. In 2D, the seed valence $\vartheta$ is a $2 \times 1$ binary array, which is associated to each seed point to indicate whether and of which type PSLs have been traced from this point. $\vartheta$ can take on four different bit combinations, i.e., empty seed $[0 \:\: 0]$ (passed by no PSL), solid seed $[1 \:\: 1]$ (passed by both major and minor PSLs) and semi-empty seed $[1 \:\: 0]$ (only passed by major PSL) or $[0 \:\: 1]$ (only passed by minor PSL). The sampling process is repeated until all valences of all possible seed points become solid $[1 \:\: 1]$. With this definition of seed valence, the sampling process is performed iteratively, by using the seed valence to characterize the state of each seed point at a specific iteration. To ensure that the generated PSLs are space-filling, the initial candidate seed points (with $\vartheta = [0 \:\: 0]$) are 
located at the vertices of a space-filling Cartesian grid (step 0 in \autoref{fig:methodOverview}). 

Seeding starts by selecting one of the candidate seed points and tracing the major and minor PSLs from it (Step 1 in \autoref{fig:methodOverview}), setting $\vartheta=[1 \:\: 1]$ at this point. Per default, the system starts with the seed point closest to the center of the bounding box of the domain, to preserve an existing plane symmetry of the stress field in the PSLs (see \autoref{fig:cantilever} and \autoref{fig:rod}). Then, all candidate seed points with $\vartheta$ not equal to $[1 \:\: 1]$ are re-classified with respect to the currently existing PSLs. To exclude candidates too close to an existing major or minor PSL, $\vartheta$ of these candidates is set to $[1 \:\: 0]$ or $[0 \:\: 1]$, respectively. If a point is classified as $[1 \:\: 0]$ or $[0 \:\: 1]$ and  closer to a minor or major PSL, respectively, its valence is set to $[1 \:\: 1]$. The distance between a point and a PSL is computed as the minimum distance between the point and any of the integration points on the PSL. Proximity is decided via a distance threshold $\varepsilon$, which also controls the density of the extracted PSLs.

\begin{figure}[t]
    \centering
    \includegraphics[width=0.98\linewidth, clip=true]{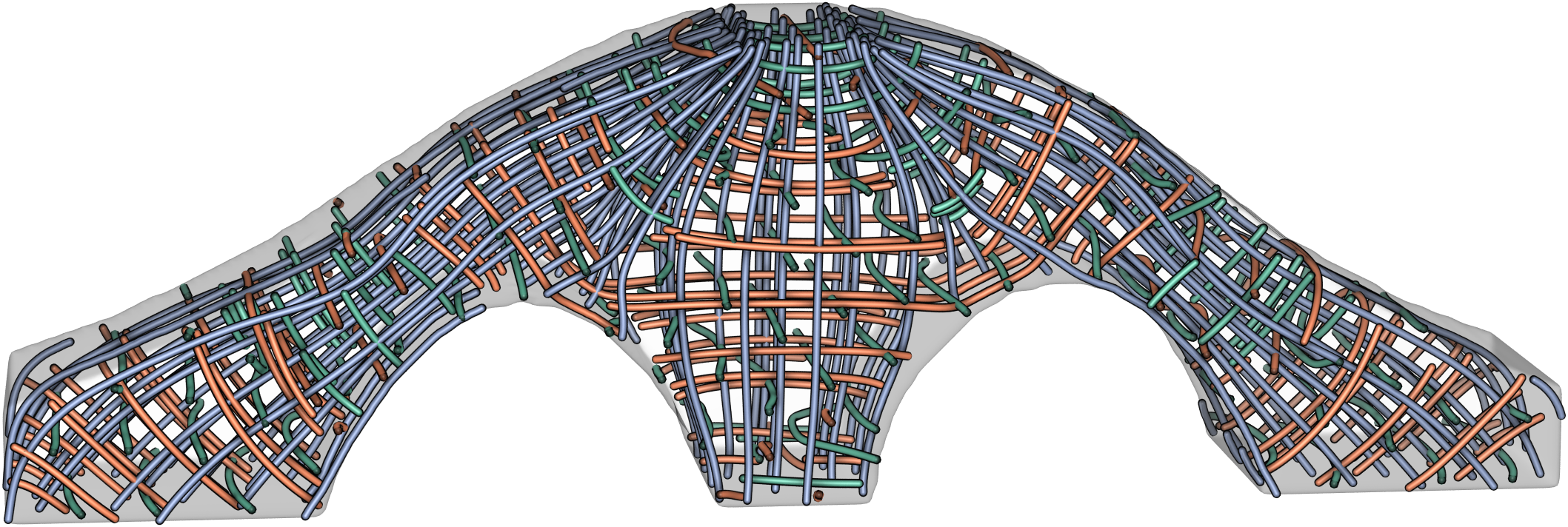} \\
    \includegraphics[width=0.98\linewidth, clip=true]{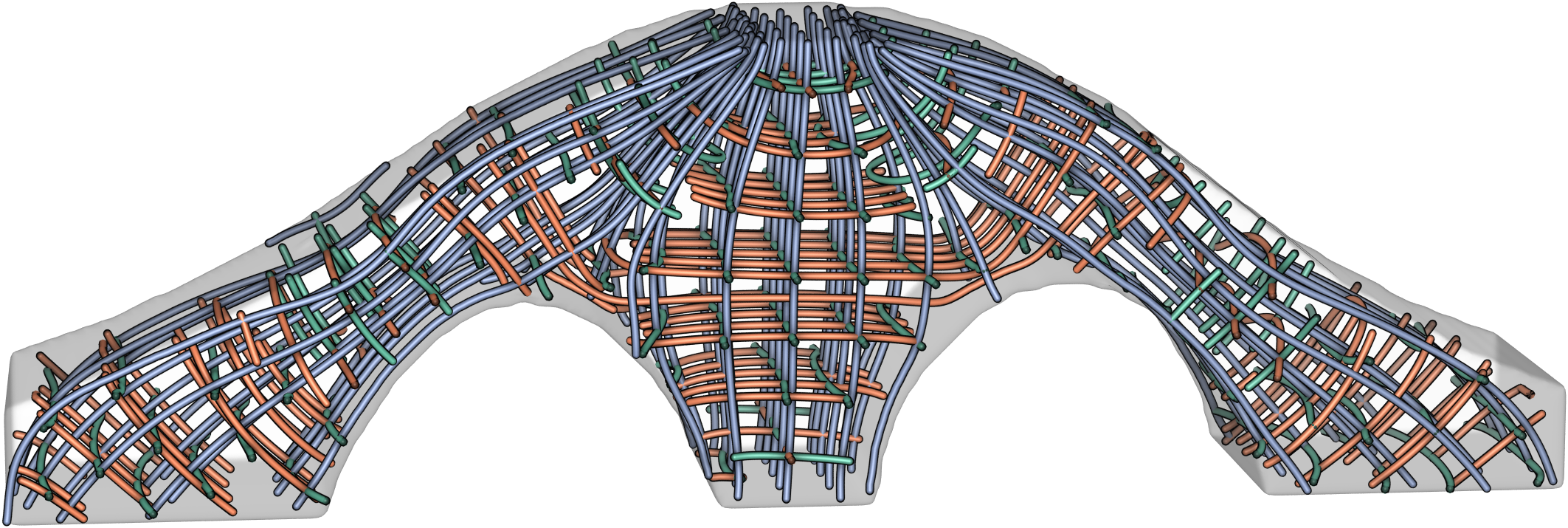} 
    \caption{PSLs in a bridge under load (see \autoref{fig:solidObjs} for the simulated load conditions). Major (ocher), medium (green) and minor (blue) PSLs generated by (top) separate seeding as proposed by \cite{Jobard1997creating} in each principal stress direction field, and (bottom) by our method. Note that since the stress field is not strictly symmetric, the PSL set shows some asymmetry.
    }
    \label{fig:comparison3D}
\end{figure}

To obtain a more regular PSL structure, each re-classified candidate point is re-located (i.e., merged) to the position of the closest integration point on the PSL causing its classification. This creates an "empty" band around the PSLs where no candidate seed point exists. The merging operation enforces that newly selected seed points lie on an existing PSL, so that the final PSL structure appears more regular and less cluttered 
(see \autoref{fig:comparison3D} for a comparison to the seeding approach by Jobard and Lefer).
By placing the initial seed point in a region deemed important, the user can specifically enforce regularity in this region. 

If the last computed PSL was a major or a minor PSL, then the next seed point is selected from the set of candidates with $\vartheta = [1 \:\: 0]$ or $[0 \:\: 1]$, respectively. Thus, we alternate the order of major and minor PSL extraction to obtain a uniform distribution of both types. Of all these, the one closest to the initial seed point is selected as the new seed point, and the respectively transverse PSL is computed. The entire procedure is then restarted until no more candidate is available (see steps 2-5 in \autoref{fig:methodOverview}).

We further consider the situation where some empty seed points may get too close (measured by $\varepsilon$) to the other type of existing PSLs after they are merged to the current PSL, e.g., the seed valence $\vartheta$ of some empty seed points become $[1 \:\: 0]$ after merging them to the newly traced major PSL. However, it can also happen that some of these merged seed points might be close to some of the existing minor PSLs, which would unavoidably cause inappropriate placement of minor PSLs in the final visualization. Given this, we identify those semi-empty seed points after merging, and compute the distances of them to the corresponding type of PSLs. If there are distances less than $\varepsilon$, the valences of these seed points are set to $[1 \:\: 1]$. By simply making $\vartheta$ a binary array with three elements referring to the major, medium and minor PSL, the proposed seeding strategy can be lifted to 3D. 

\subsection{PSL LoD Structure} \label{SubSec:LoD}
To change the density of the generated PSLs, the seeding process can simply be re-run with an appropriately set distance threshold $\varepsilon$. The larger this threshold is, the less PSLs are extracted. However, the different sets of PSLs that are generated for different thresholds are not nested, i.e., the PSLs at a coarser representation with lower PSL density are not a subset of the PSLs at a representation with higher density. Therefore, in an exploration session where the user interactively selects different PSL LoDs, there are abrupt changes when transitioning from one level to another. To avoid this, we propose to generate a nested PSL hierarchy. 

The basic idea underlying the construction of a nested hierarchy is to let the PSLs at a level with higher PSL density 'grow out' sequentially from the PSLs at a lower density level. As a side effect, this enables saving computations by progressively computing a new level from the previous coarser level. For a given set of PSLs that have been generated with distance threshold $\varepsilon_0$, the refined set of PSLs according to a distance threshold $\varepsilon_1 < \varepsilon_0$ is computed as follows: Firstly, the candidate seed points are reset to their initial positions. Secondly, the candidate seed points are merged to the existing PSLs according to $\varepsilon_1$, to create ``empty'' bands around the existing PSLs. The valences are updated accordingly to $[1 \:\: 0]$, $[0 \:\: 1]$ or $[1 \:\: 1]$ depending on the types of PSL they are merged to. After this, some non-solid seeds are left, because $\varepsilon_0$ is larger than $\varepsilon_1$. With these seeds the seeding is subsequently performed, including the iteration of seed point selection, PSL computation, and re-classification as described in \autoref{SubSec:seeding}. 

\begin{figure}[t]
    \centering
    \includegraphics[width=0.24\linewidth]{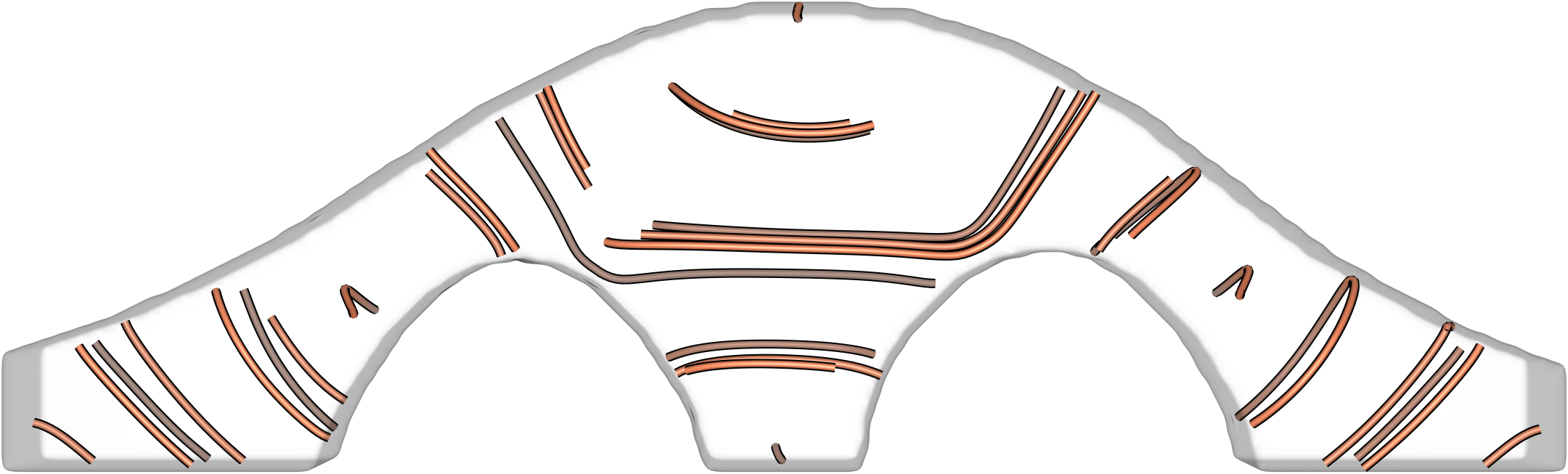}
    \includegraphics[width=0.24\linewidth]{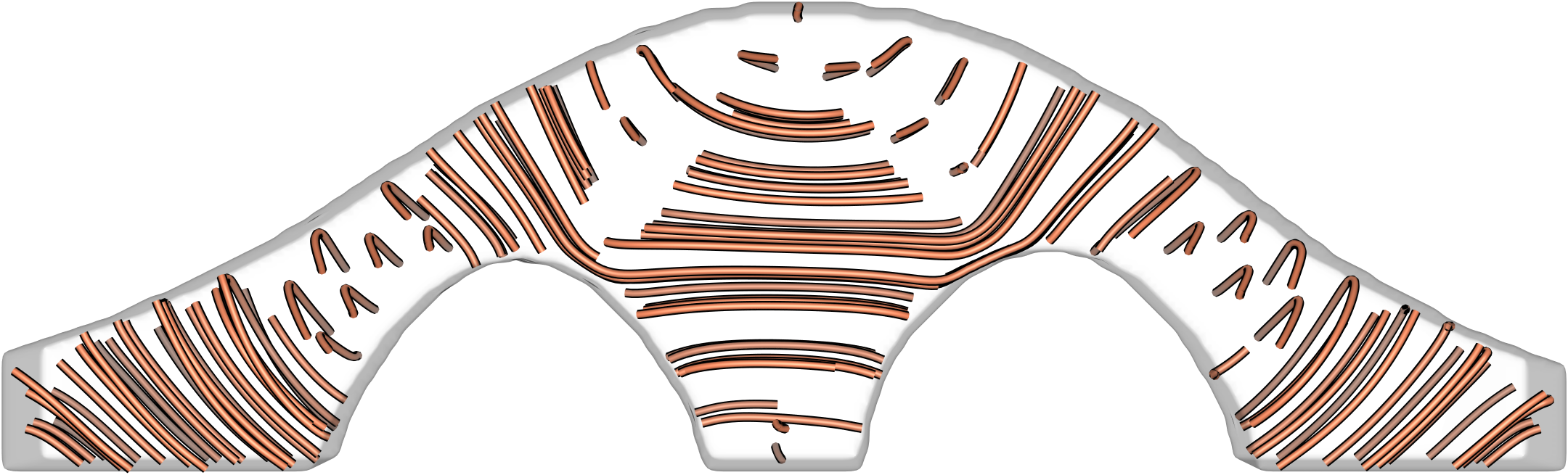} 
    \includegraphics[width=0.24\linewidth]{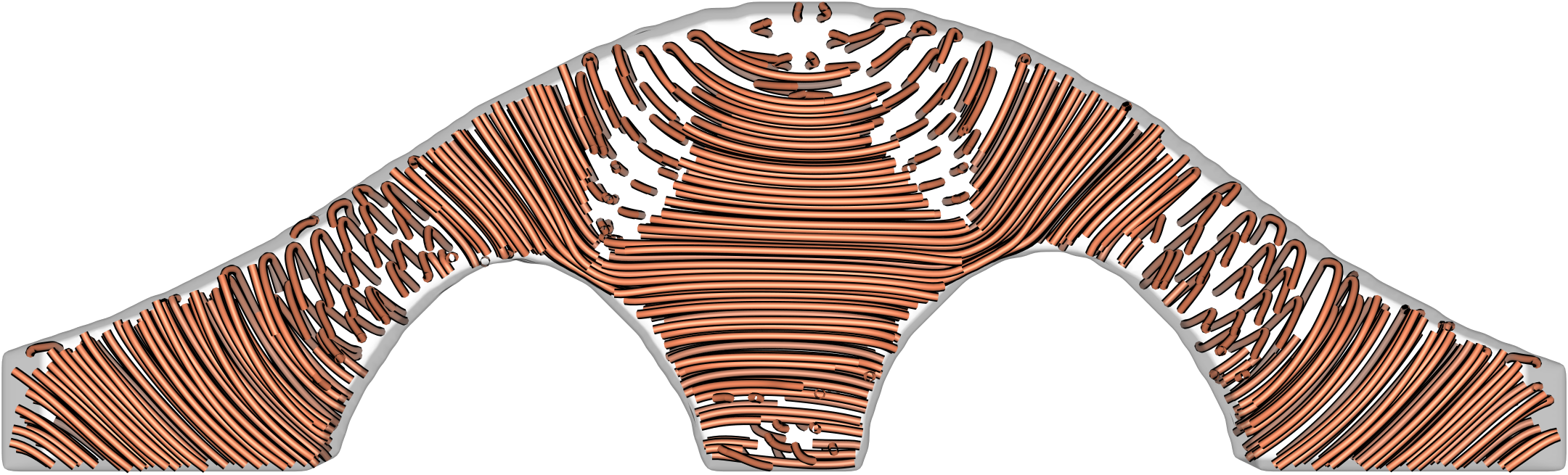} 
    \includegraphics[width=0.24\linewidth]{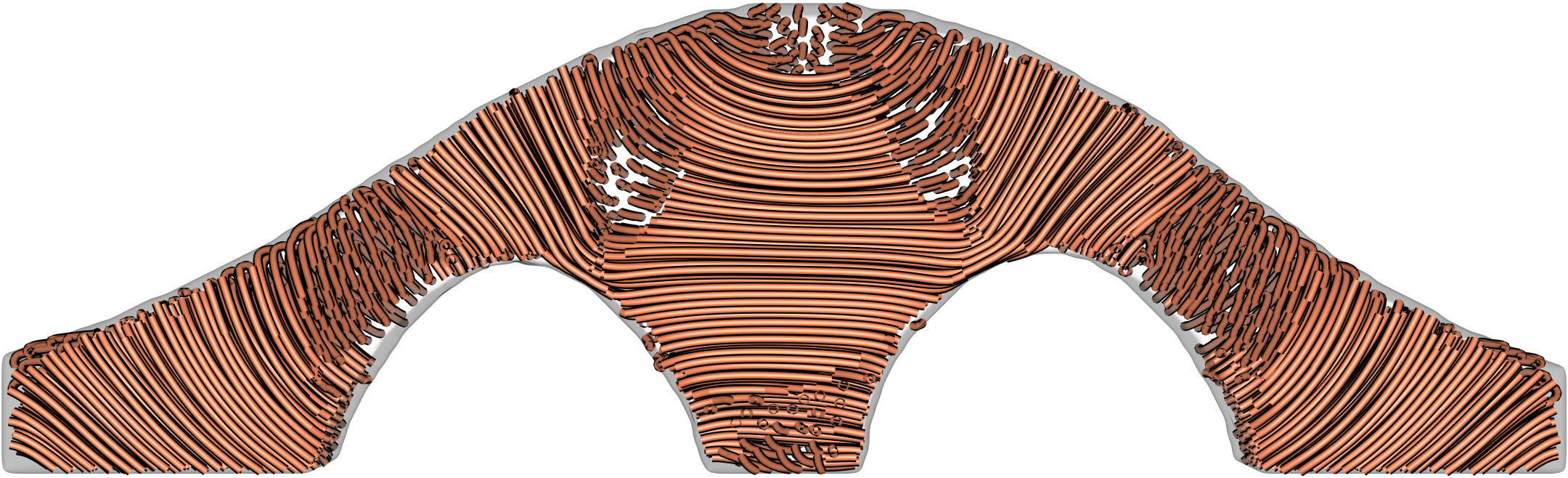} \\
    \includegraphics[width=0.24\linewidth]{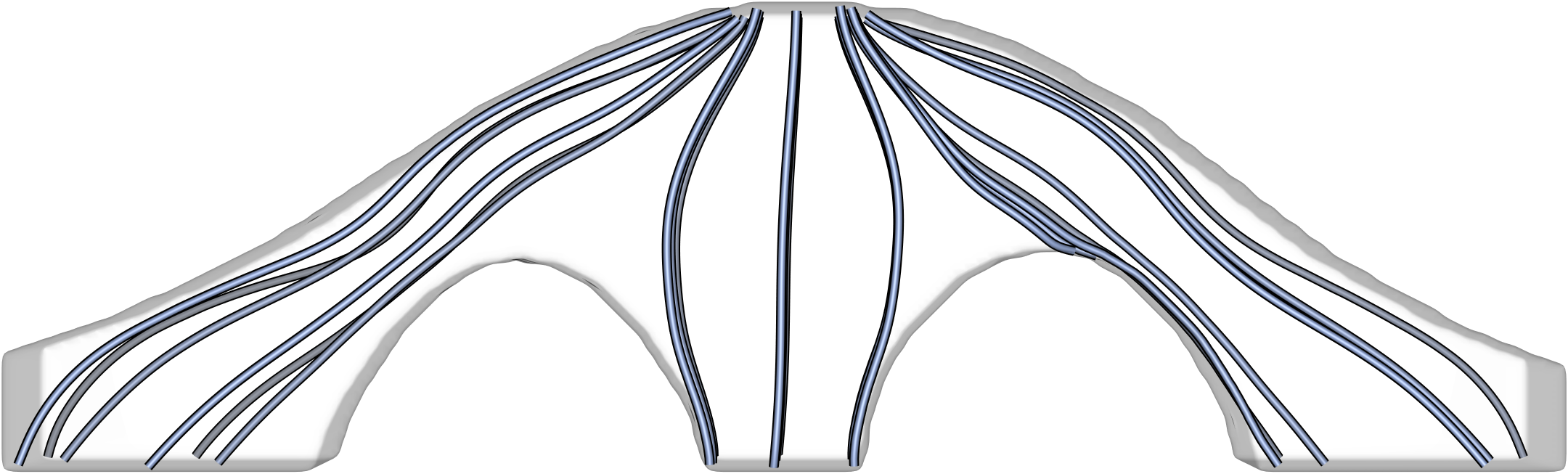}
    \includegraphics[width=0.24\linewidth]{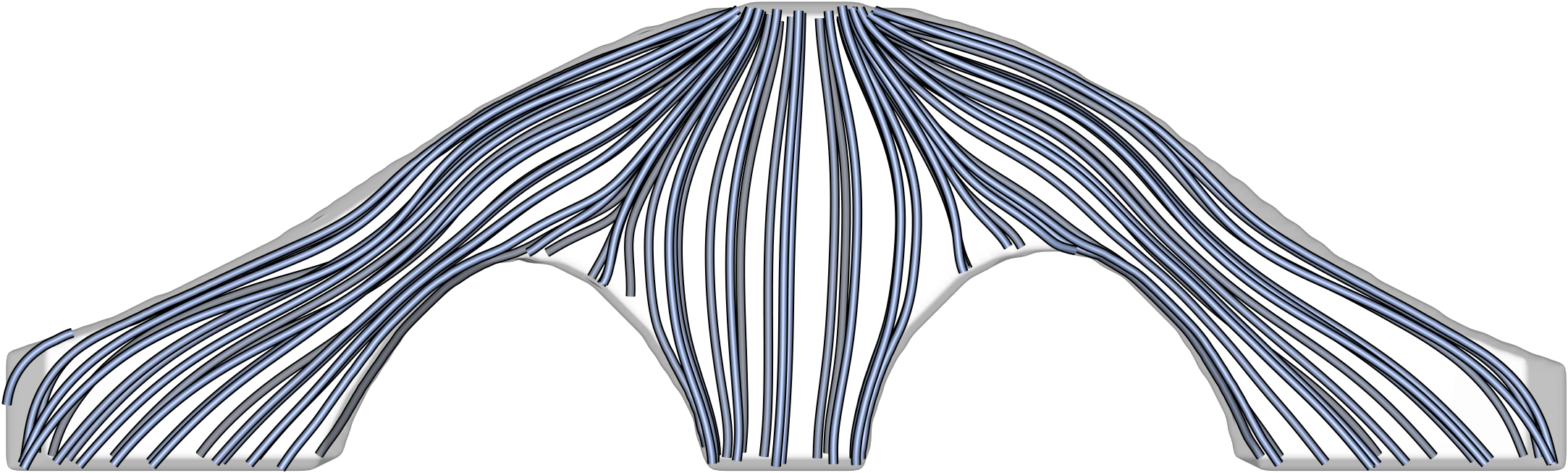} 
    \includegraphics[width=0.24\linewidth]{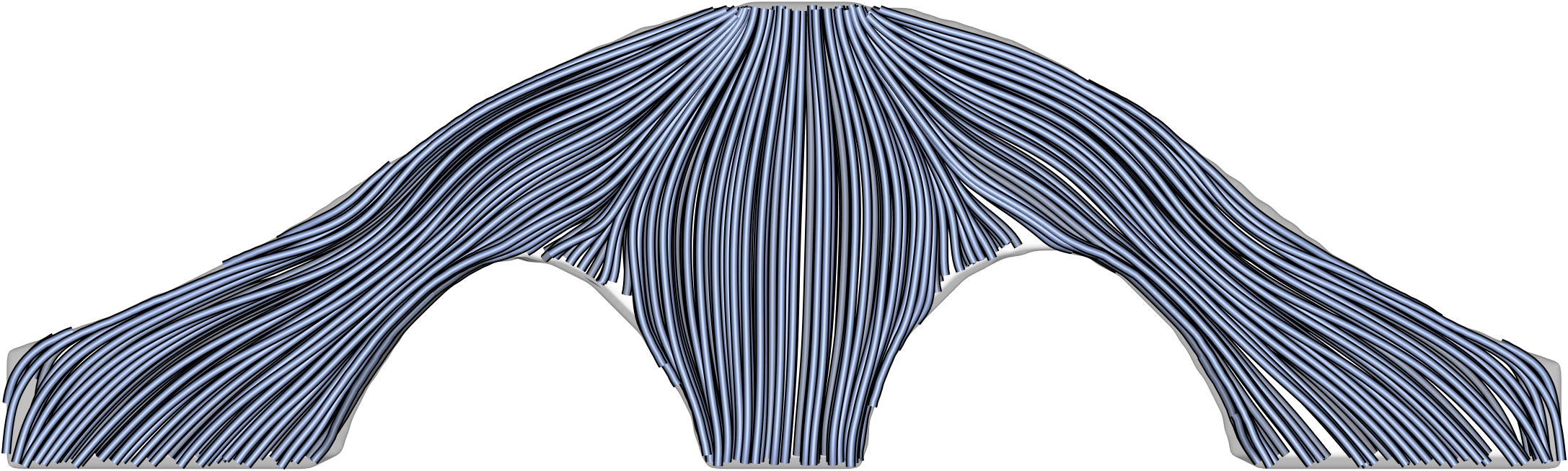} 
    \includegraphics[width=0.24\linewidth]{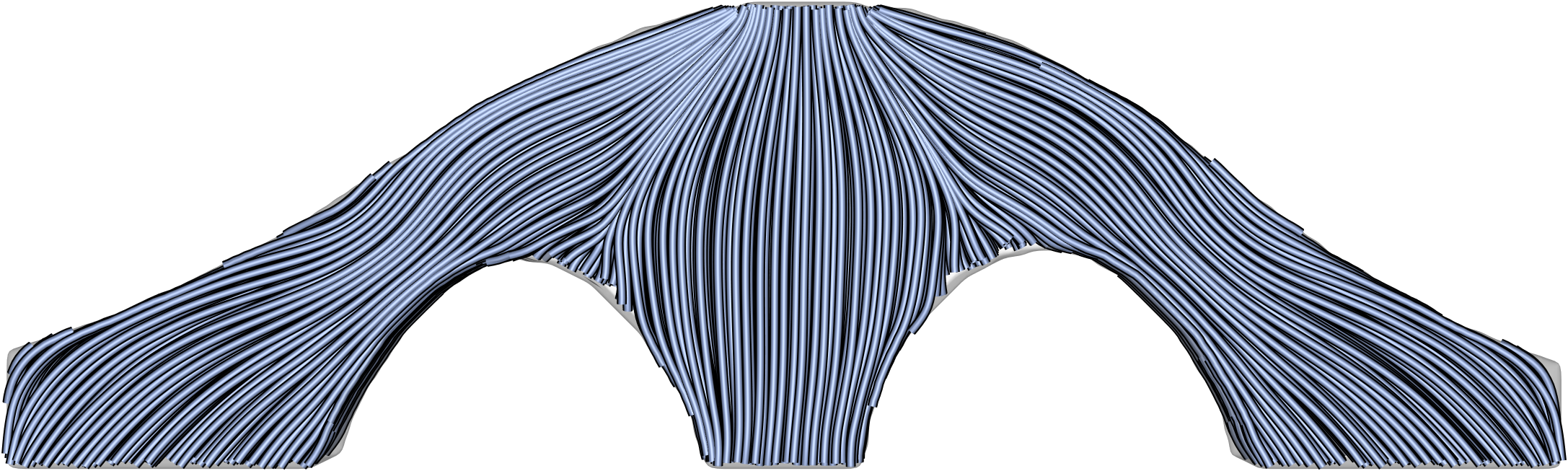} \\
    \includegraphics[width=0.96\linewidth]{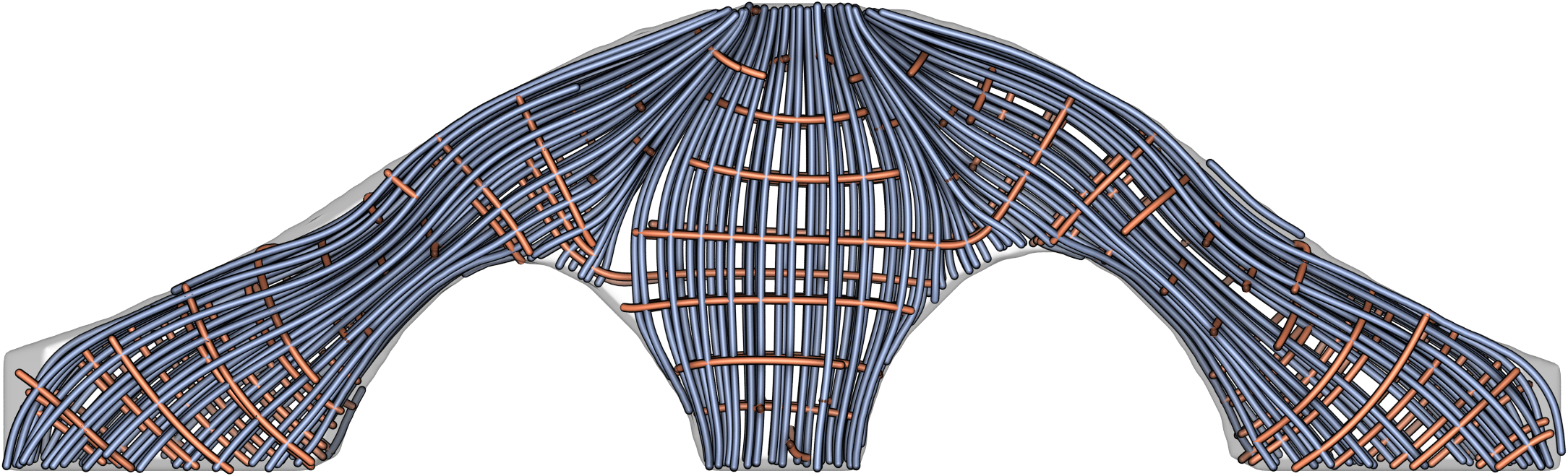}
    \caption{PSL LoD hierarchy. Top: The major and minor PSLs at different LoDs, computed separately for each level. Bottom: Simultaneous extraction of the PSL structure using level $L2$ (context) for the major and $L3$ (focus) for the minor PSLs.
    }
    \label{fig:Sketch_MultiResolutionHierarchy}
\end{figure}

To generate a full LoD PSL hierarchy, the user defines the minimum distance threshold $\varepsilon$ and the number $M$ of levels to construct. Then, the distance thresholds of each level are computed as $2^{(M-k)}\varepsilon, \:\: k=1:M$ from coarse to fine, and the hierarchy is constructed progressively from the coarsest resolution level (see 1st and 2nd rows in \autoref{fig:Sketch_MultiResolutionHierarchy}). To compute a PSL structure with different types of PSLs at different LoDs, the distance thresholds for each PSL type are first selected by the user, and then the multi-type LoD is computed by alternatively considering the different PSL types with their respective distances.

\subsection{Ribbon-based Stress Visualization} \label{SubSec:PSLgeo}
\begin{figure}[t]
    \centering
    \includegraphics[width=0.98\linewidth]{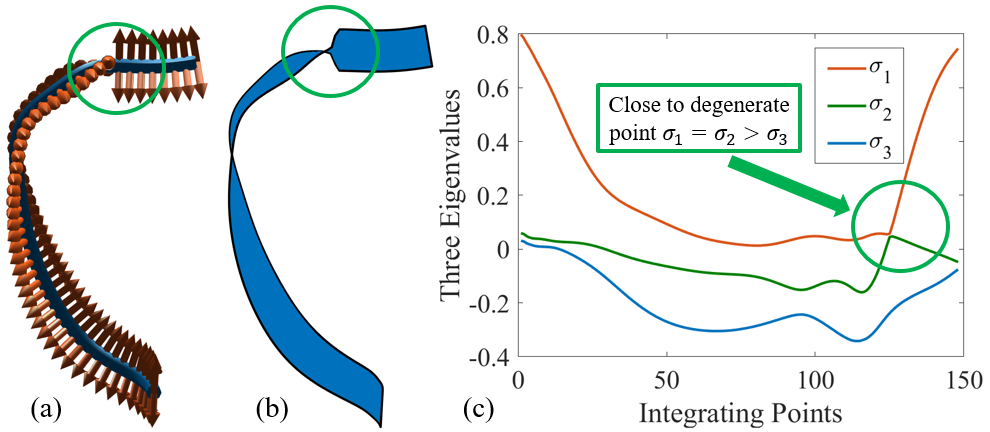}
    \caption{When a PSL goes through a degenerate point (a), the ribbon-shaped geometry shows a sudden twist (b). (c) Behaviour of the eigenvalues along the ribbon's center PSL, from which the ribbon's direction and orientation is determined.}
    \label{fig:ribbon_schematics_degenerate}
\end{figure}

Instead of rendering lines, the user can select a PSL type (i.e., major, medium, minor) and visualize ribbon-shaped geometry~\cite{Streamribbons} that is centered at the PSLs of the selected type and twists according to the direction of another stress type
(see \autoref{fig:ribbon_schematics_degenerate}~a,b). At each integration point along a PSL of the selected type, two lines with adjustable length are traced forward and backward along the other direction. The lines' endpoints at subsequent integration points are connected to form a ribbon. It is worth noting that the constructed ribbons don't coincide with streamsurfaces that are integrated from a PSL along one other stress direction. As shown by Raith et al.~\cite{Raith2018Tensor}, such surface might not even exist, i.e., when integrating from two points on the same PSL over a certain length along another stress direction, the two endpoints are not lying on a PSL in general. The mapping of two principal stress directions to a ribbon geometry is conceptually similar to the well-known hyperstreamlines~\cite{Delmarcelle1993Visualizing}, i.e., a mapping of two principal stress directions to a tube centered at the PSL along the third direction. 

We let the user select a visualization using ribbons to convey changes in the assignment of the eigenvector directions to the type of PSL in the vicinity of degenerate points. When a ribbon is formed as described, flips often occur in the vicinity of a degenerate point (see \autoref{fig:ribbon_schematics_degenerate}~(c)). This is because the two directions can exchange their classification as major, medium, and minor, since this depends only on their position in the sorted sequence of eigenvalues. Thus, ribbons provide an additional visual cue to indicate topological changes of the PSLs in the vicinity of degenerate points. 



\begin{figure}[t]
    \centering
    \includegraphics[width=0.98\linewidth]{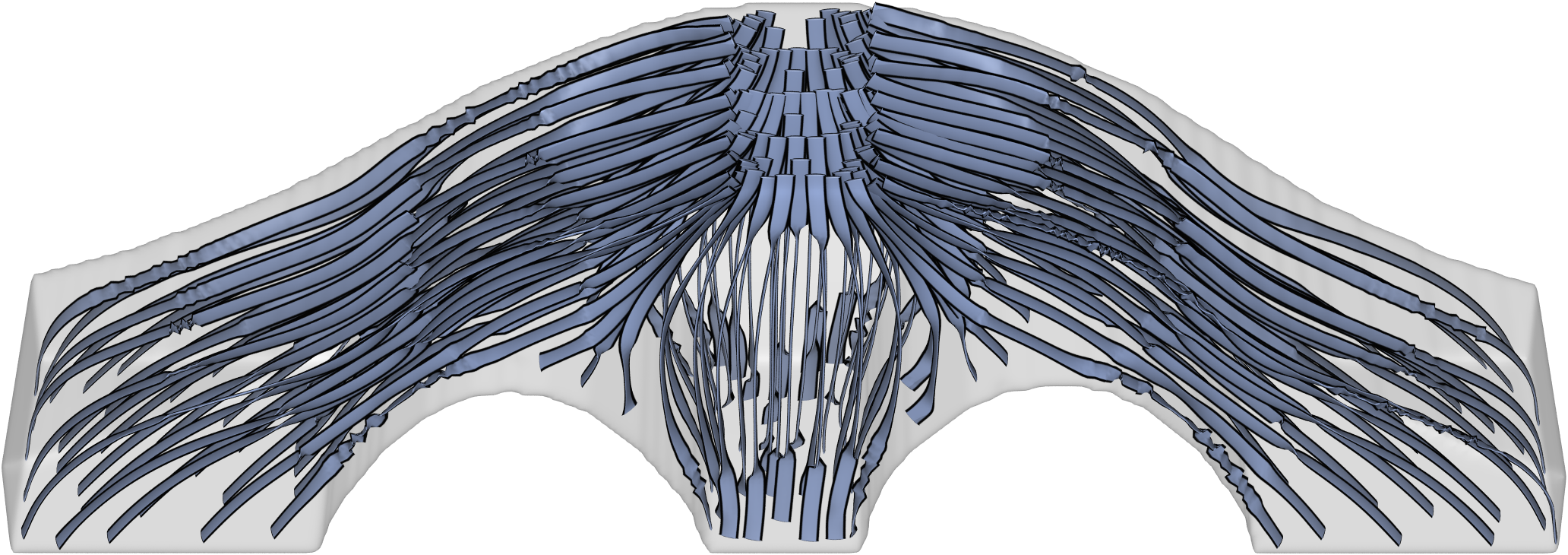}\\
    \includegraphics[width=0.98\linewidth]{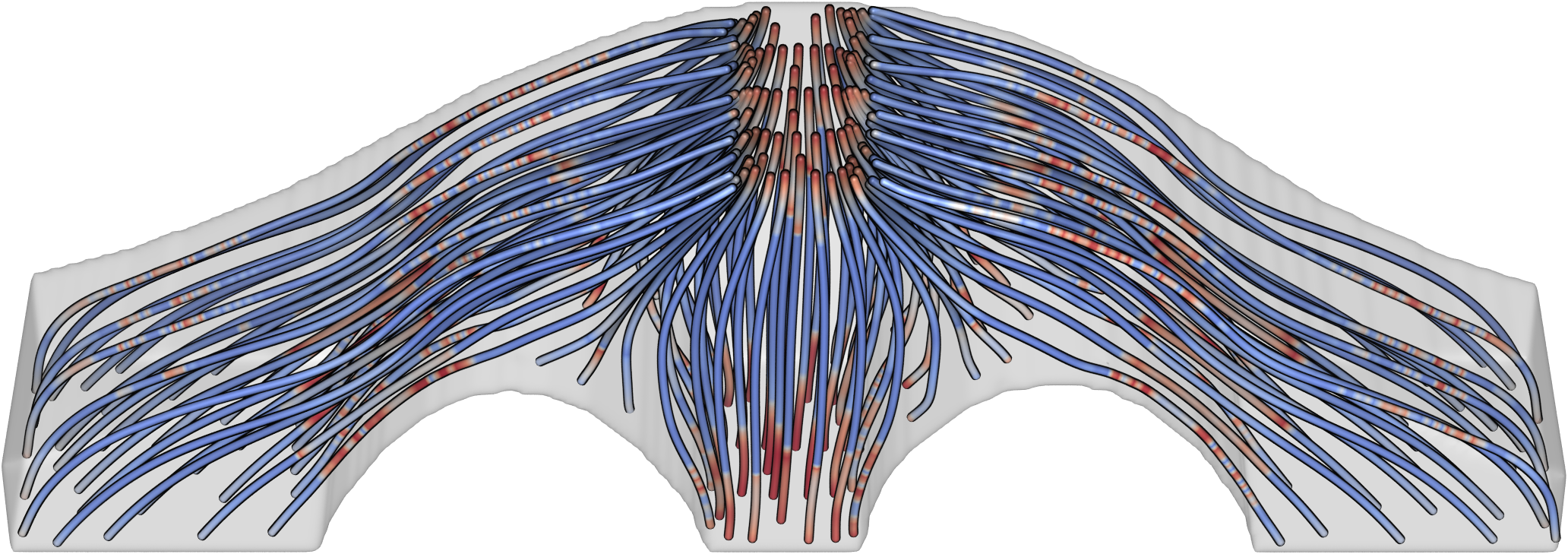}\\
    \includegraphics[width=0.98\linewidth]{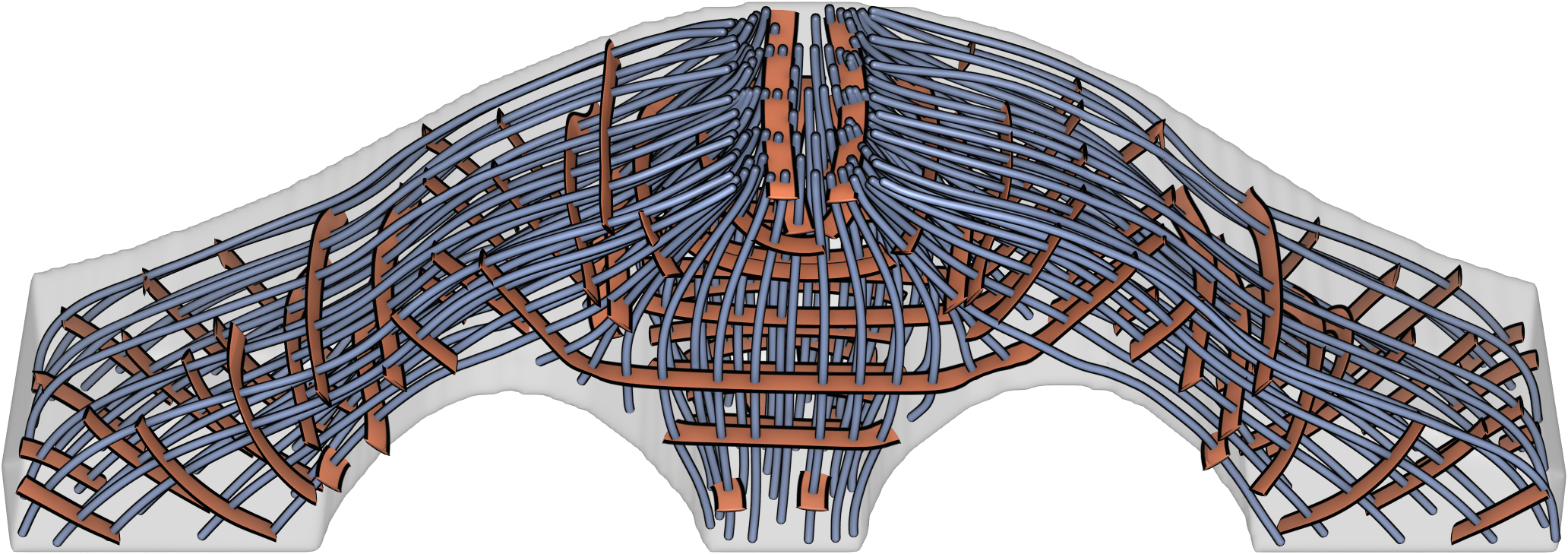}
    \caption{Top: Ribbons are aligned along the minor PSLs and twist according to the medium principal stress direction. Middle: Minor PSLs with degeneracy measure mapped from blue (low) to red (high). Bottom: A visualization using lines for minor PSLs and ribbons for major PSLs.
    }
    \label{fig:SmoothingComparison}
\end{figure}

\autoref{fig:SmoothingComparison} compares the options to visualize principal stress directions via ribbons and lines, and combine them into a single visualization. As can be seen, twists in the ribbon geometry effectively hint to regions where degenerate points might exist. For lines, 3D-TSV can map the degeneracy measure introduced in Sec.~\ref{Sec:PrelimWork} to color. An interesting observation is that high degeneracy and flips thereof frequently occur close to the object boundaries when Cartesian simulation meshes are used. These flips occur due to the well-known inaccuracies at curved boundaries that are represented by hexahedral simulation elements in a Cartesian grid. 

\section{System Implementation} \label{Sec:sysImp}
To implement the communication between the C++ visualization frontend and the MatLab extraction backend, the messaging library ZeroMQ is utilized, which can be used for communication over a wide variety of protocols, like TCP/IP. 3D-TSV relies on the request-reply pattern implemented in ZeroMQ, where the frontend issues a new request to the backend when the user changes simulation settings in the graphical user interface, and the backend sends back a reply as soon as the simulation is finished in order to notify the frontend of the availability of new data.

The reason why we turned to MatLab instead of C++ for the implementation of the backend is, on the one hand, that the sampling method is an inherently sequential algorithm. Thus, it cannot benefit significantly from multi-threaded PSL tracing or GPU parallelization. On the other hand, MatLab is widely spread in engineering, where most of our collaborators regarding stress visualization come from, and the engineers tend to use mainstream commercial software they are already familiar with to finish the design iteration quickly. In this case, they can run the MatLab backend independently without any complicated compilation and setup process. To this end, we also provide a slim MatLab visualization implementation, which can provide users a fast and easy way to explore the stress field, while discarding some more complex hardware-accelerated features from the C++ frontend, like depth cues or ambient occlusion effects. It is worth noting that also the rendering frontend can be used standalone, by reading trajectories from a file specifying the exchange format regarding PSL type and LoD representation.

\subsection{Numerical PSL Integration}
3D-TSV is designed to support the visualization of PSLs in solids discretized by hexahedral grids, where the stress tensors are given at the grid vertices. When computing PSLs in Cartesian grids, component-wise trilinear interpolation of the tensors is used during numerical line integration. In deformed hexahedral cells, tensor interpolation is performed via inverse distance weighting~\cite{shepard1968two}.

To integrate PSLs in Cartesian grids, the system provides fixed-step integration schemes with user adjustable stepsize of at least half the cell diameter. In deformed hexahedral grids, a different approach is taken since the size of the simulation elements can vary, and with a constant stepsize the risk increases that multiple cells smaller than this size are missed in one single integration step. 
To reduce this risk, the integration stepsize is automatically adapted to the size (i.e., the length of the shortest edge) of the cell at the current integration point $P_i$. These values are pre-computed and stored per cell. In each integration step, the size $s$ of the current cell is read and multiplied by a user selected scaling factor $\delta_s$. $\delta_s$ can be made smaller than 1 to obtain more accurate PSLs. With the stepsize $s \cdot \delta_s$, the PSL is integrated from the current point $P_i$ in cell $e_i$ to the new point $P_{i+1}$. Then, the integration process is restarted with $P_{i+1}$ and the cell $e_{i+1}$ containing $P_{i+1}$.

\begin{figure}[t]
    \centering
    \includegraphics[width=0.98\linewidth,trim=0.0cm 0.0cm 0.0cm 0.0cm, clip=true]{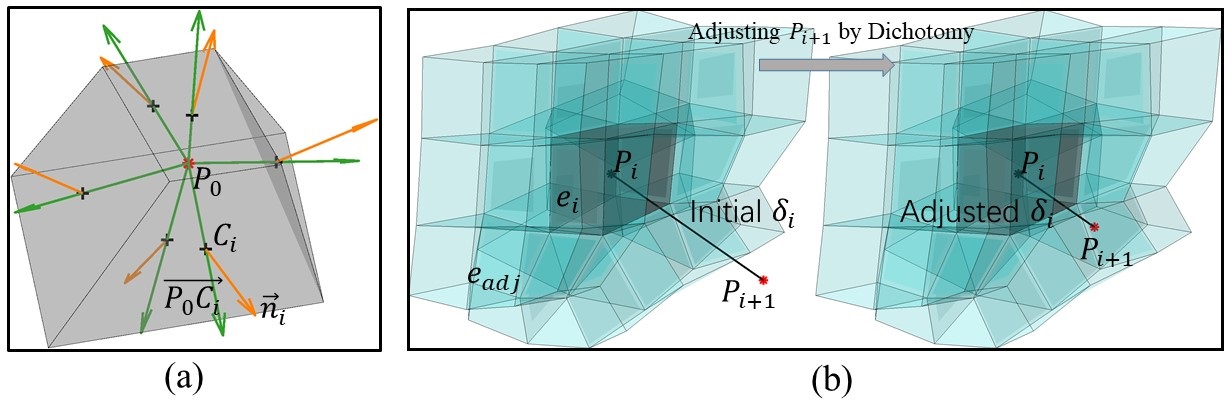}
    \caption{(a) Quantities required to test whether a point $P_0$ (red $*$) is located in a hexahedral cell. Black "$+$" and orange arrows indicate centers $C_i$ and out-facing normals $\vec{n}_i, \:\: i\in \{1,\cdots, 6\}$ of the six cell faces. Green arrows indicate the directional vectors $\vec{P_0C_i}, \:\: i\in \{1,\cdots, 6\}$ that are used. (b) Point re-location is subsequently performed until the next integration point $P_{i+1}$ is within the same cell $e_i$ (grey cube) as the current point $P_i$, or is within one of the cells $e_{adj}$ (cyan cubes) adjacent to $e_i$. 
    }
    \label{fig:TracingUnstructuredMesh}
\end{figure}

To find $e_{i+1}$, it is first tested whether $P_{i+1}$ is still contained in $e_i$. The following in-out criterion is used to test whether a point is located in a hexahedral cell: Given a hexahedral element with the centers and out-facing normal of its 6 faces $C_i$ and $\vec{n}_i, \:\: i\in \{1,\cdots, 6\}$. Any point $P_0$ in the interior or on the boundary of the element satisfies $\max(\arccos(\vec{P_0C_i},\vec{n}_i)) \leq \frac{\pi}{2}, \:\: i\in \{1,\cdots, 6\}$, see~\autoref{fig:TracingUnstructuredMesh}a. In practice, the criterion is slightly relaxed to $\max(\arccos(\vec{P_0C_i},V_i)) \leq \frac{91\pi}{180}, \:\: i\in \{1,\cdots, 6\}$, to account for non-planar cell faces, i.e., a slight variation of the normal vectors across the faces.

If $e_i$ does not contain $P_{i+1}$, the cell $e_{i+1}$ needs to be determined. To this end, we further test whether $P_{i+1}$ lies in any of the adjacent cells $e_{adj}$ of $e_i$. For each cell, the set of adjacent cells as well as the adjacency type, i.e., face-, edge-, and vertex-adjacency, is pre-computed and stored. In case $P_{i+1}$ is not within $e_i$ or $e_{adj}$, we scale down the stepsize via a dichotomy strategy, i.e., $P_{i+1}=(P_{i+1}+P_i)/2$, until $P_{i+1}$ is located in $e_i$ or it's adjacent cells $e_{adj}$.

\begin{figure}[t]
    \centering
    \includegraphics[width=0.95\linewidth,trim=0.0cm 0.0cm 0.0cm 0.0cm, clip=true]{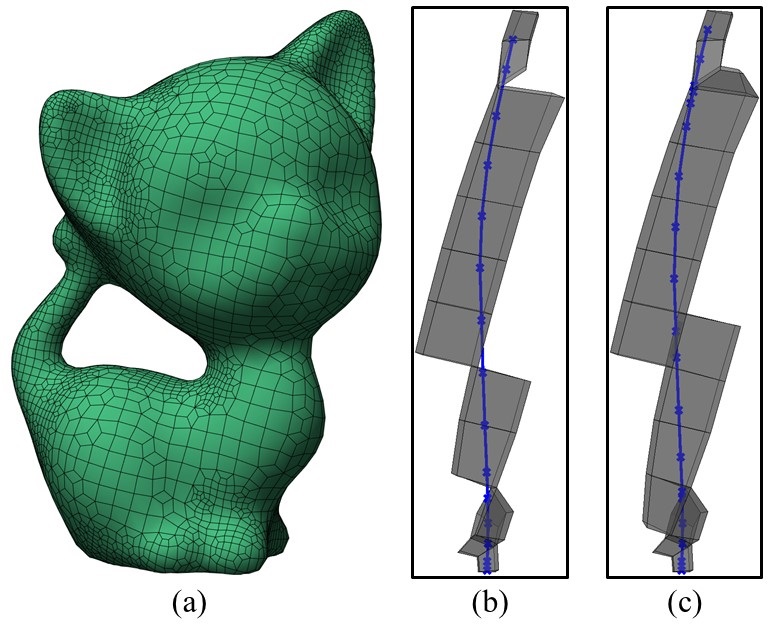}
    \caption{(a) The deformed hexahedral simulation mesh. (b) A PSL (blue trajectory) in the simulated stress field. It is ensured that every next integration point is in the previous cell or in a cell adjacent to the previous cell. (c) Same as (b), but now every next integration point is in a face-adjacent cell.}
    \label{fig:continuityMetric}
\end{figure}

In the case where $e_i$ and $e_{i+1}$ are connected by a single edge or vertex, it may still happen that cells are skipped when going from $P_{i}$ to $P_{i+1}$. 
In this situation, stepsize refinement is performed multiple times until 
the cell $e_{i+1}$ shares a face with $e_i$ or is below a user-selected threshold. The latter situation is encountered when the PSL goes through a cell vertex or edge, so that face-adjacency cannot be determined. 
In \autoref{fig:continuityMetric}, for the given mesh two PSLs that have been extracted without and with additional stepsize refinement are compared. As can be seen, cells that would be skipped when using only face-to-face adjacency are now determined and considered in the integration.

\subsection{Rendering}
The line and ribbon primitives are rendered in a stylized fashion similar to the techniques by Zöckler et al.~\cite{IllumStreamlines}, Stoll et al.~\cite{StylizedLines} and Mattausch et al.~\cite{IlluminatedStreamlines}, using default colors, halos and depth cues as shown in the first three images in \autoref{fig:teaser}. Focus PSLs and contextual ribbons are rendered in ocher and blue, respectively. The base color is modulated using Blinn-Phong shading \cite{BlinnPhong,IllumStreamlines}, which assumes a point light source at the world space position of the viewer (i.e., a head light).

The user can interactively change the color mapping---also separately for each PSL type---and can in particular switch to a mapping of some scalar quantity to color, as indicated in the last image in \autoref{fig:teaser} using the scalar von Mises stress measure. The scalar values are issued via the backend as per-vertex attributes. 
The standard color scheme we use for the different principal stress directions (blue, green, ocher) is the `3-class Set2' transfer function from ColorBrewer\footnote{\url{https://colorbrewer2.org/\#type=qualitative&scheme=Set2&n=3}}. It is colorblind safe and print friendly.

For enhanced depth perception, depth cues are added, i.e., with increasing distance to the camera, fragments are increasingly desaturated. A translucent simulation mesh outline hull can be rendered together with the stress field data in order to hint at the extents of the simulation domain. 

\subsection{3D-TSV Settings} \label{SubSec:DoG}

\begin{figure*}[ht]
    \centering
    \includegraphics[width=0.98\linewidth,trim=0.0cm 0.0cm 0.0cm 0.0cm, clip=true]{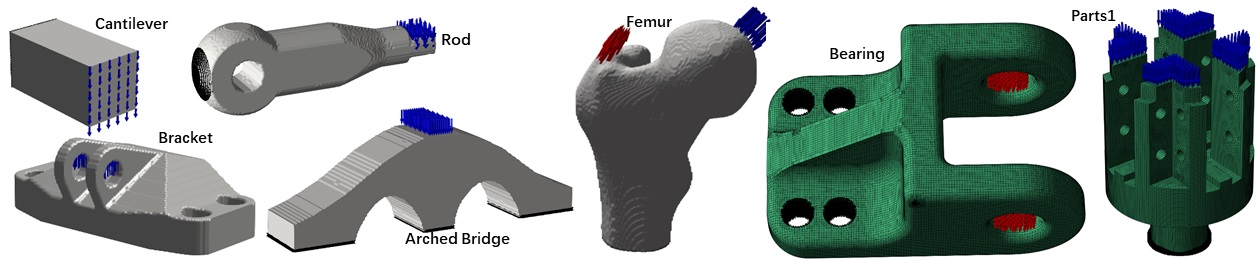}
    \caption{The solid objects used in this work and the applied external loads. Red and blue arrows indicate the loading positions and directions, black regions indicate fixed boundaries. A finite-element-based elasticity analysis has been used to compute the stress field for each model under the predicted loads. The unstructured hexahedral meshes `Parts' and 'Bearing' are courtesy of~\cite{li2012all} and \cite{DBLP:journals/cgf/GaoSP19}, respectively. All other meshes are Cartesian meshes. `Arched Bridge' and 'Rod' are courtesy of~\cite{arora2019volumetric} and \cite{DBLP:journals/cgf/GaoSP19}, respectively. All simulated stress fields are made publicly available.}
    \label{fig:solidObjs}
\end{figure*}

\begin{table}[t]
\small
\begin{center}
\scalebox{0.85}{
\begin{tabular}{ccccccc}
	\hline
	Data Set & \#Cells & \#Seeds & $\varepsilon / D_{0}$ & $M$ & \#PSLs & Time (s)\\ 
    \hline
    Cantilever & 250K & 2K & 1/5 & 1 & 85 & 0.4\\
    Rod & 536K & 18K & 1/5 & 1 & 174 & 2.1 \\    
    Femur & 696K & 10K & 1/18 & 3 & 823 & 9.0\\
    Bracket & 650K & 9K & 1/12 & 3 & 293 & 5.4\\
    Bearing & 189K & 55K & 1/18 & 3 & 1,364 & 33.4\\    
    Parts1 & 253K & 46K & 1/20 & 3 & 1,557 & 27.9\\
    \hline
\end{tabular}}
\end{center}
\caption{Model and performance statistics. $D_{0}$ is the length of the shortest dimension of the bounding box of the stress field.}
\label{tab:expCond}
\vspace{-0.2cm}
\end{table}

3D-TSV provides a number of parameters that can be changed by the user to control the generation of PSLs. These parameters include the merging threshold $\varepsilon$ and the number of levels $M$ introduced in \autoref{SubSec:seeding} and \autoref{SubSec:LoD}, respectively. Another set of parameters enables a user-guided interaction with the PSL distribution, including sliders for controlling the LoD resolution of major, medium and minor PSLs. In addition, the user can select the two PSL types that are used to generate ribbons. Via a drop-down menu, the user can select a scalar stress measures that are mapped to PSL color using a transfer function. The backend provides different stress components, such as the principal stress amplitudes, von Mises stress, and the six Cartesian stress components.

\section{Results}
In all of our experiments, PSL generation is performed on the CPU, i.e., a workstation running Ubuntu 20.04 with an AMD Ryzen 9 3900X @3.80GHz CPU and 32GB RAM. Rendering is done on an NVIDIA RTX 2070 SUPER GPU with 8GB of on-chip memory. The rendering times are always below 10 milliseconds. The data sets we use in our experiments are shown in \autoref{fig:solidObjs}. The stress fields are simulated by a finite element method (FEM), using the solid objects under the shown load conditions. \autoref{tab:expCond} lists the numbers of simulation elements of each of the data sets, the seed points that are used to generate the PSLs, the number of generated PSLs, and the time required for PSL generation. 

 

\begin{figure}[t]
    \centering
    \includegraphics[width=0.48\linewidth]{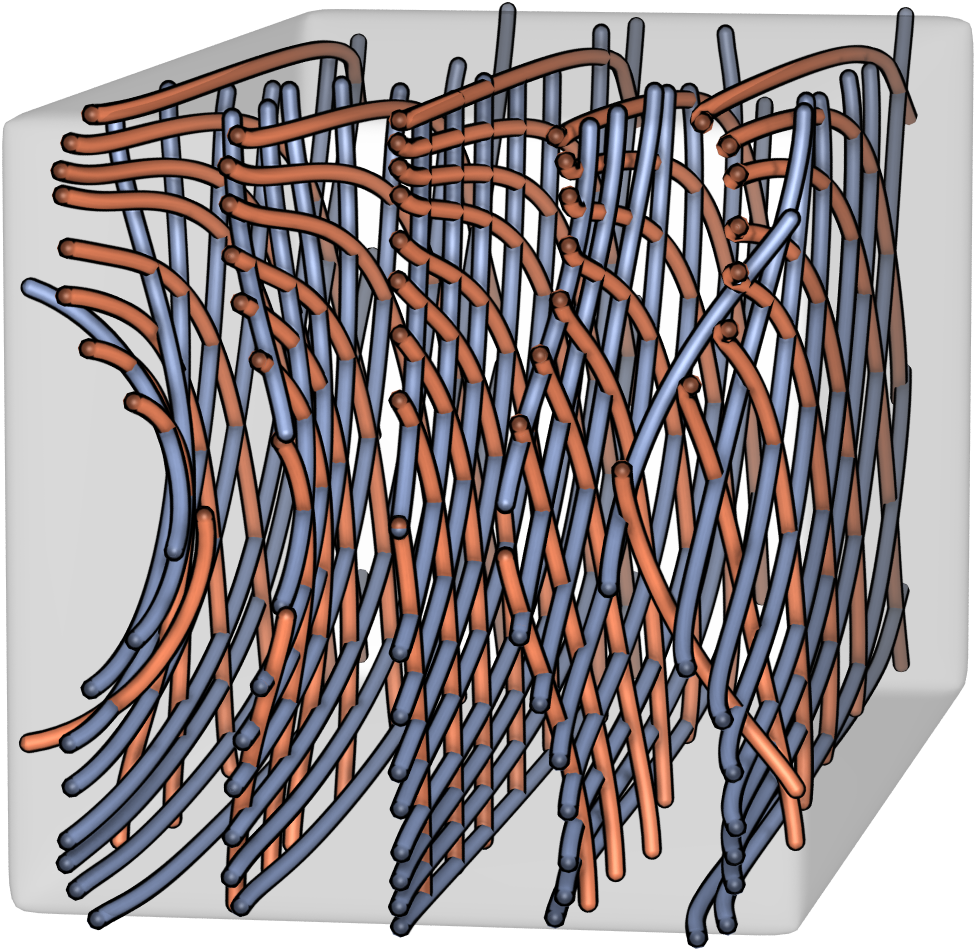}
    \includegraphics[width=0.48\linewidth]{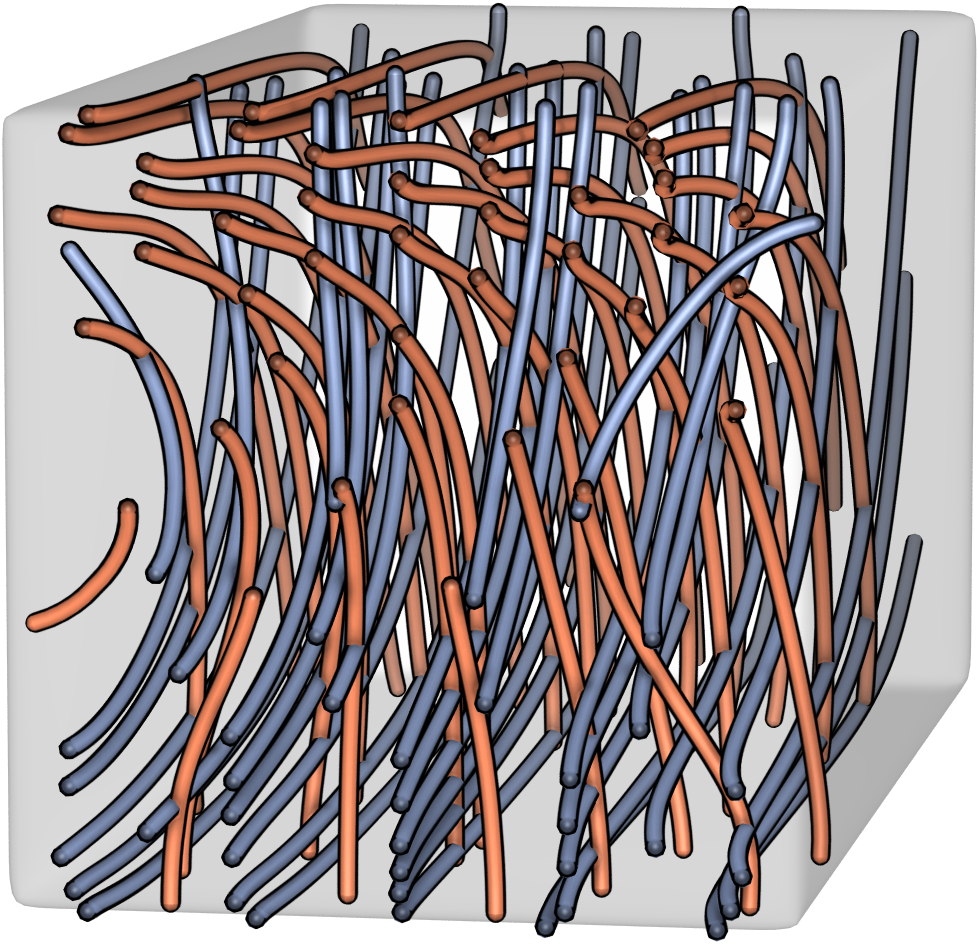}
    \caption{PSLs in the `Cantilever' stress field. PSLs by the proposed seeding strategy (left) and evenly spaced streamline seeding (right).}
    \label{fig:cantilever}
\end{figure}

For the three models 'Bridge', 'Cantilever' and 'Rod', we demonstrate the improvements of the proposed seeding strategy over evenly spaced streamline seeding. 3D-TSV is used to visually analyze the stress fields in 'Femur' and 'Bracket'. These two data sets that are frequently seen in structural design and optimization~\cite{Wu2018TVCG}. Finally, we consider the two mechanical parts 'Bearing' and 'Parts1' to demonstrate the application of 3D-TSV to unstructured hexahedral simulation meshes.

\begin{figure}[ht]
    \centering
    \includegraphics[width=0.98\linewidth]{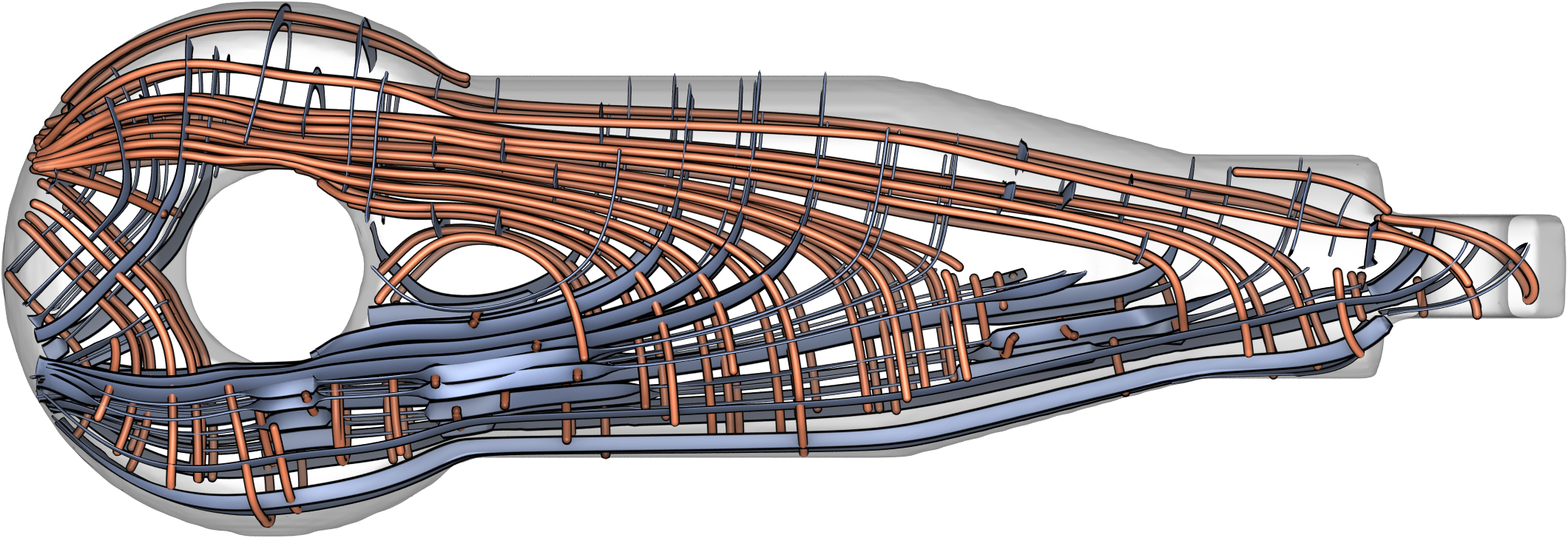} \\
    \includegraphics[height=4.5cm]{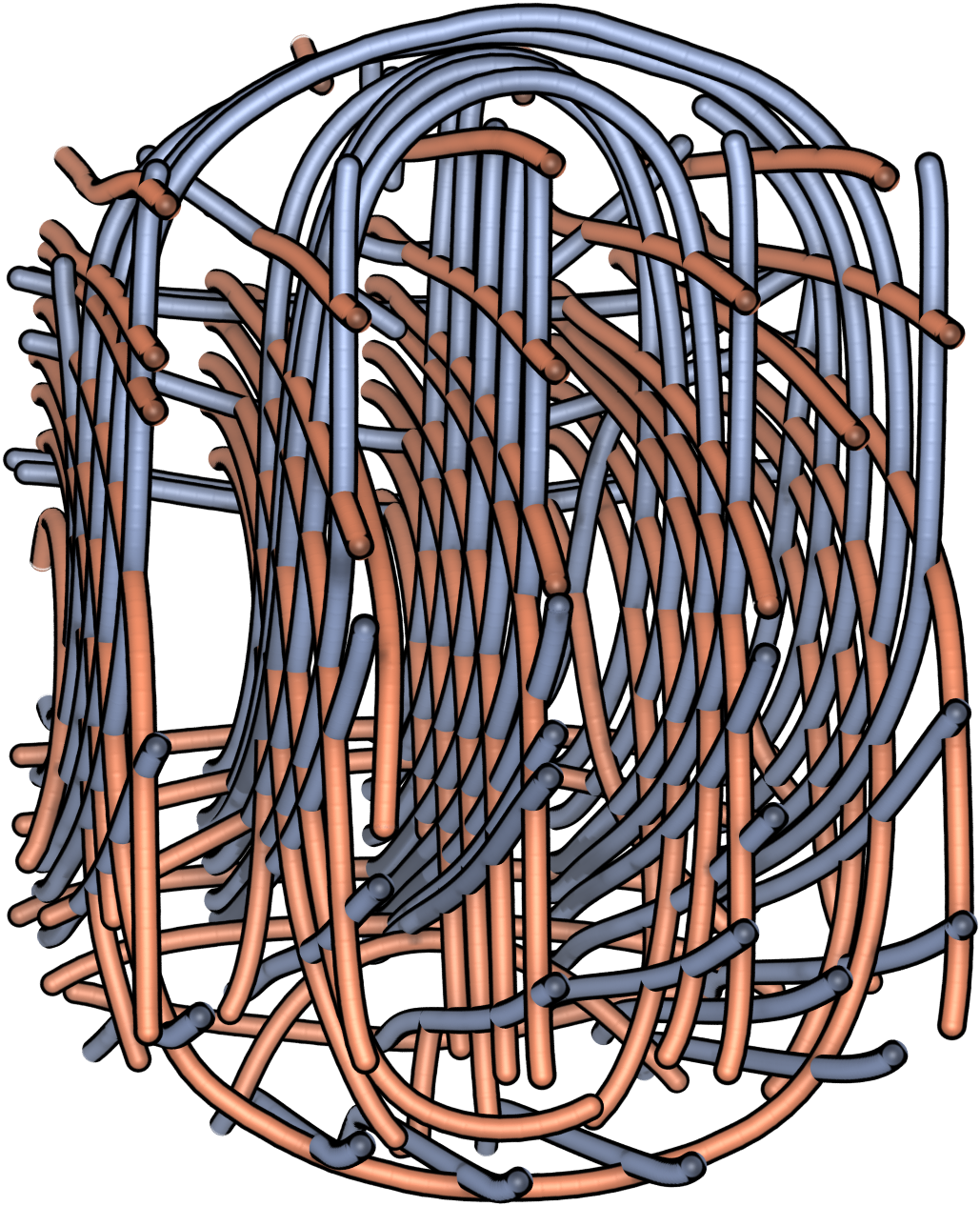}
    \includegraphics[height=4.5cm]{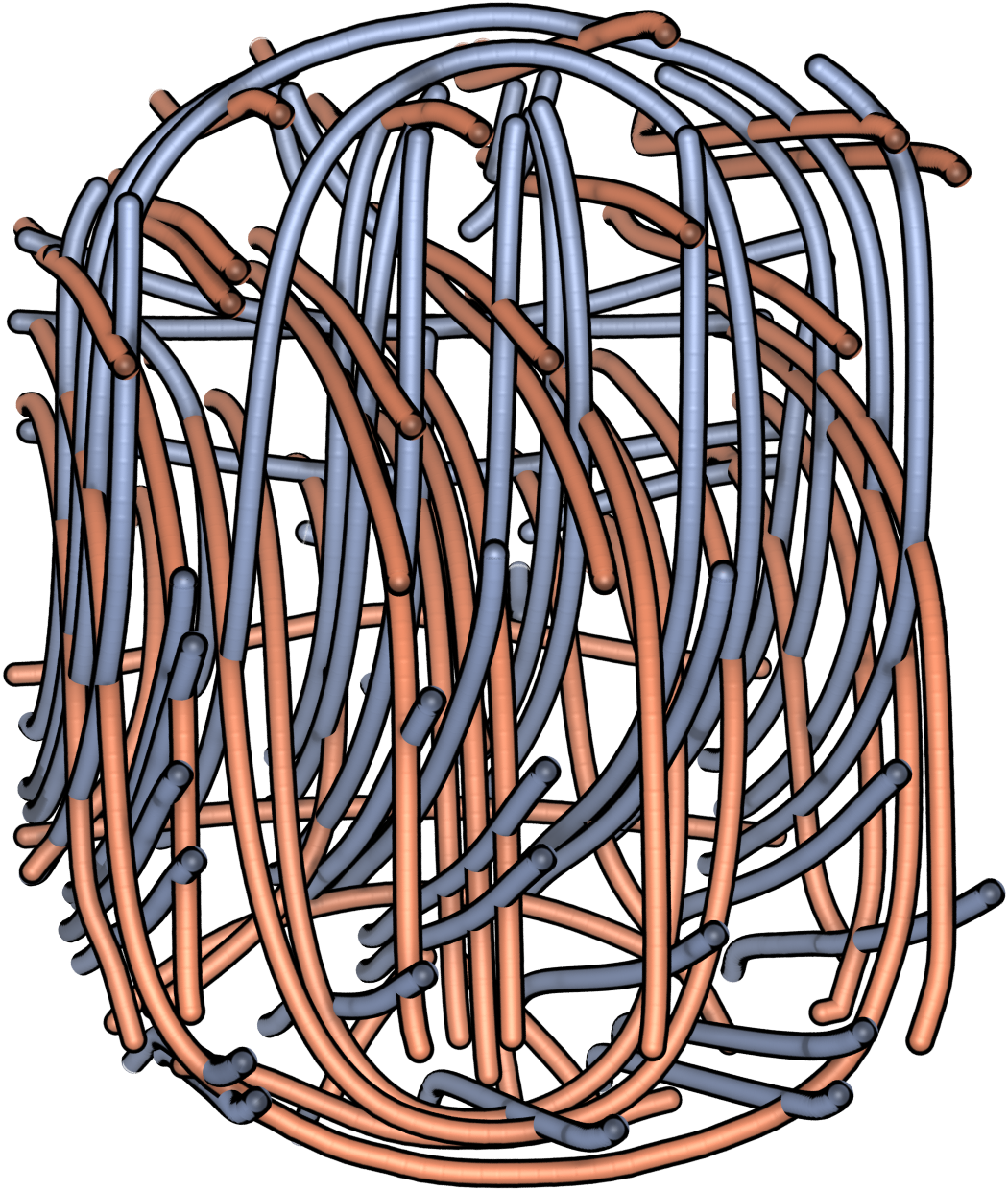} \\    
    \caption{Top: PSLs showing the principal stress directions in `Rod'. Bottom: PSLs in `Rod' from a different view. Left: PSLs computed by 3D-TSV. Right: PSLs computed via evenly spaced seeding as proposed by~\cite{Jobard1997creating}.}
    \label{fig:rod}
\end{figure}

Figs.~\ref{fig:cantilever} and~\ref{fig:rod} emphasize the improvements by the proposed seeding strategy regarding the regularity of the extracted set of PSLs. 3D-TSV generates a fairly uniform space-filling PSL structure, which, in particular, maintains the symmetry of the stress field in 'Cantilever'. Evenly spaced streamline seeding, on the other hand, generates a far less regular design which introduces severe visual clutter. 

The visualization also highlights the importance of showing different PSL types simultaneously. In the analyzed tensor field, the signs of the eigenvalues along the major and minor PSLs are mostly positive and negative, respectively. This means that the major PSLs are mainly under tension and the minor PSLs mainly under compression. Thus, either of both effects could be shown by visualizing one PSL type, but not both.  

\begin{figure}[ht]
    \centering
    \includegraphics[width=0.9\linewidth]{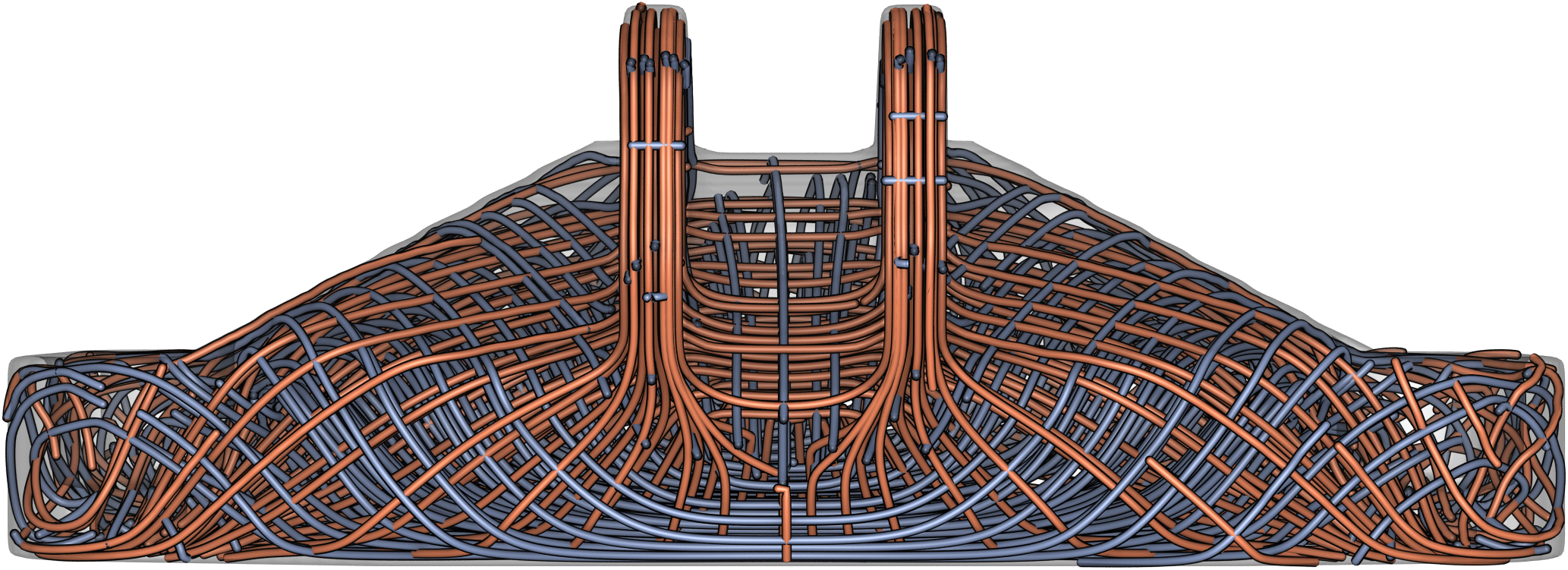}\\
    \includegraphics[width=0.9\linewidth]{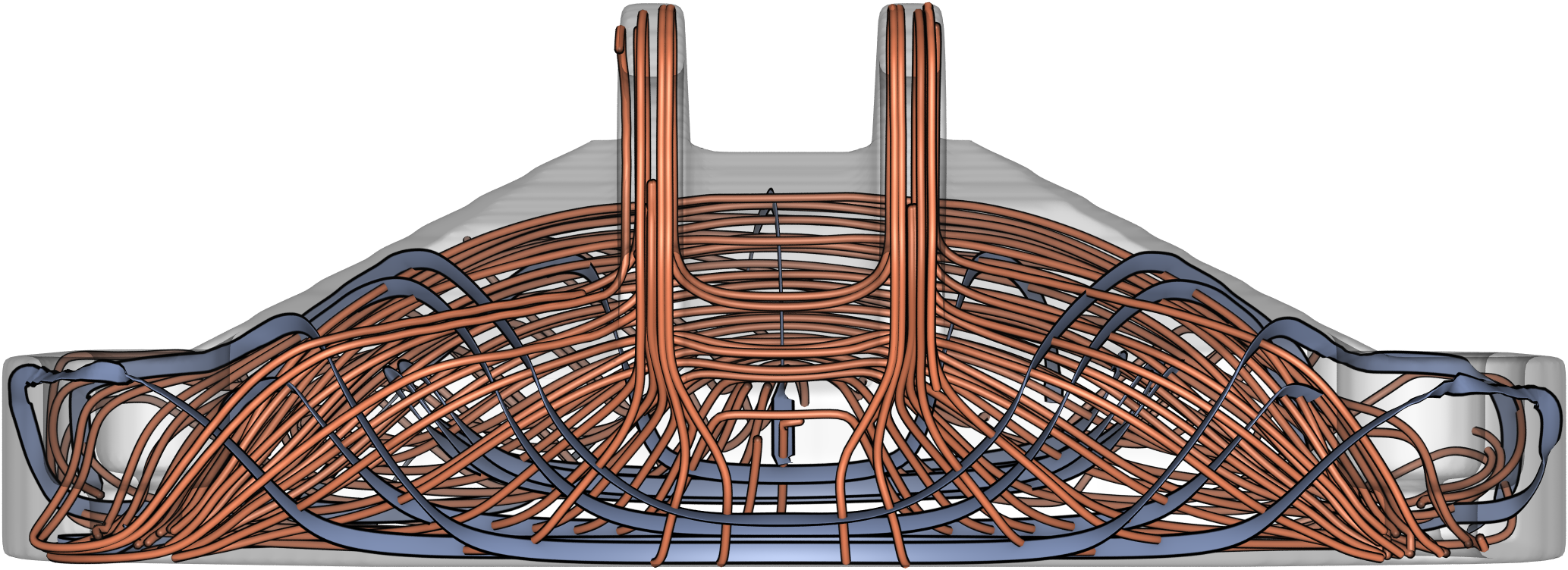}
    \caption{Stress field in `Bracket'. Top: PSLs at the finest level (according to \autoref{tab:expCond}). Bottom: major / minor PSLs at the third ($L3$) / first ($L1$) level of detail.}
    \label{fig:bracket}
\end{figure}

\autoref{fig:bracket} (top) shows the space-filling PSLs in the stress field in the interior of 'Bracket'. From the boundary condition in \autoref{fig:solidObjs}, we see that the structure is mainly under tension. Thus, we choose to show the major PSLs at the higher level of detail ($L2$) and the minor PSLs at lower level $L1$ (see \autoref{fig:bracket} (bottom)). The minor PSLs are shown via ribbons, with the medium principal stress direction indicating the twist. This enables a fine granular analysis of the major principal stress directions, and simultaneously provide a coarse representation of the other principal directions. A similar setting has been selected to visualize the stress directions in 'Femur' (see \autoref{fig:teaser}).

\begin{figure}[ht]
    \centering
    \includegraphics[width=0.98\linewidth]{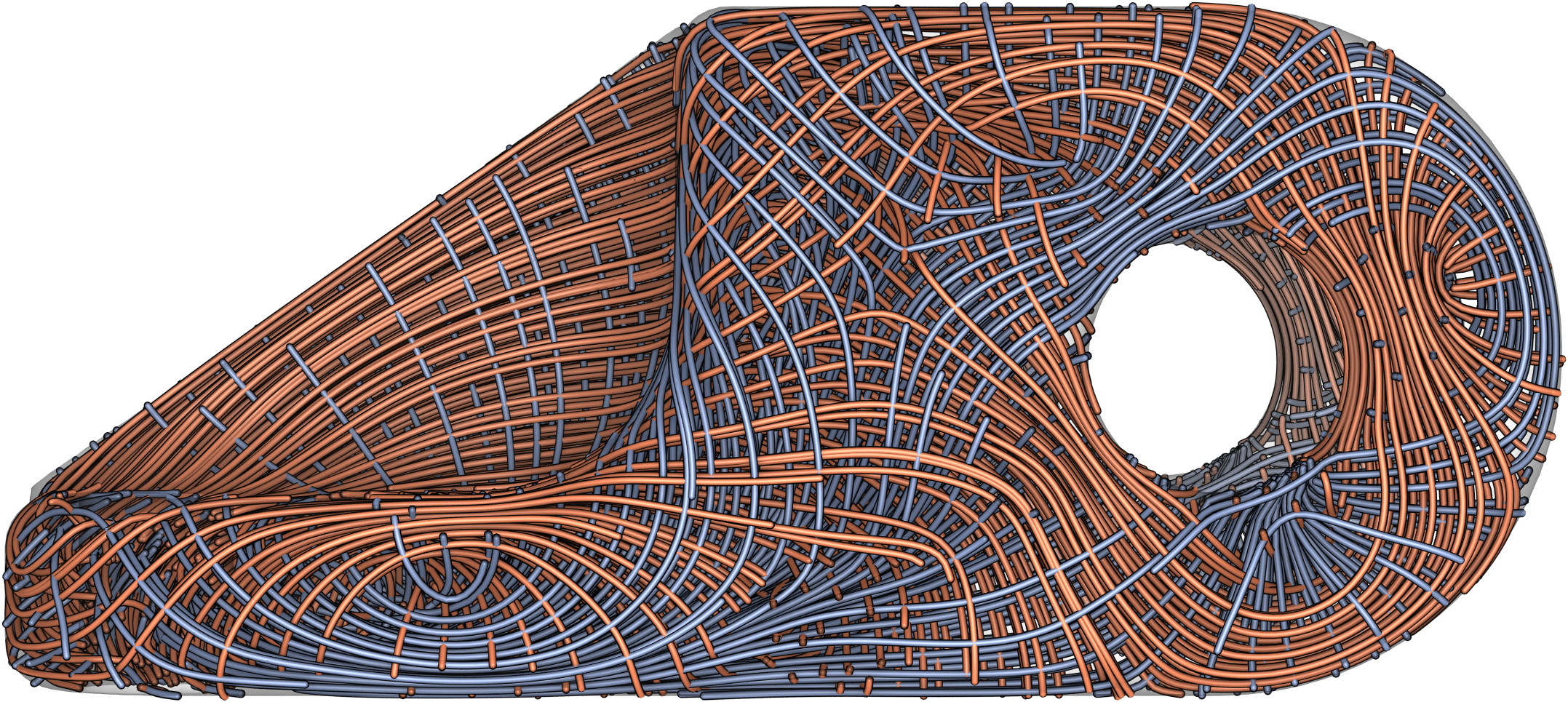} \\
    \includegraphics[width=0.98\linewidth]{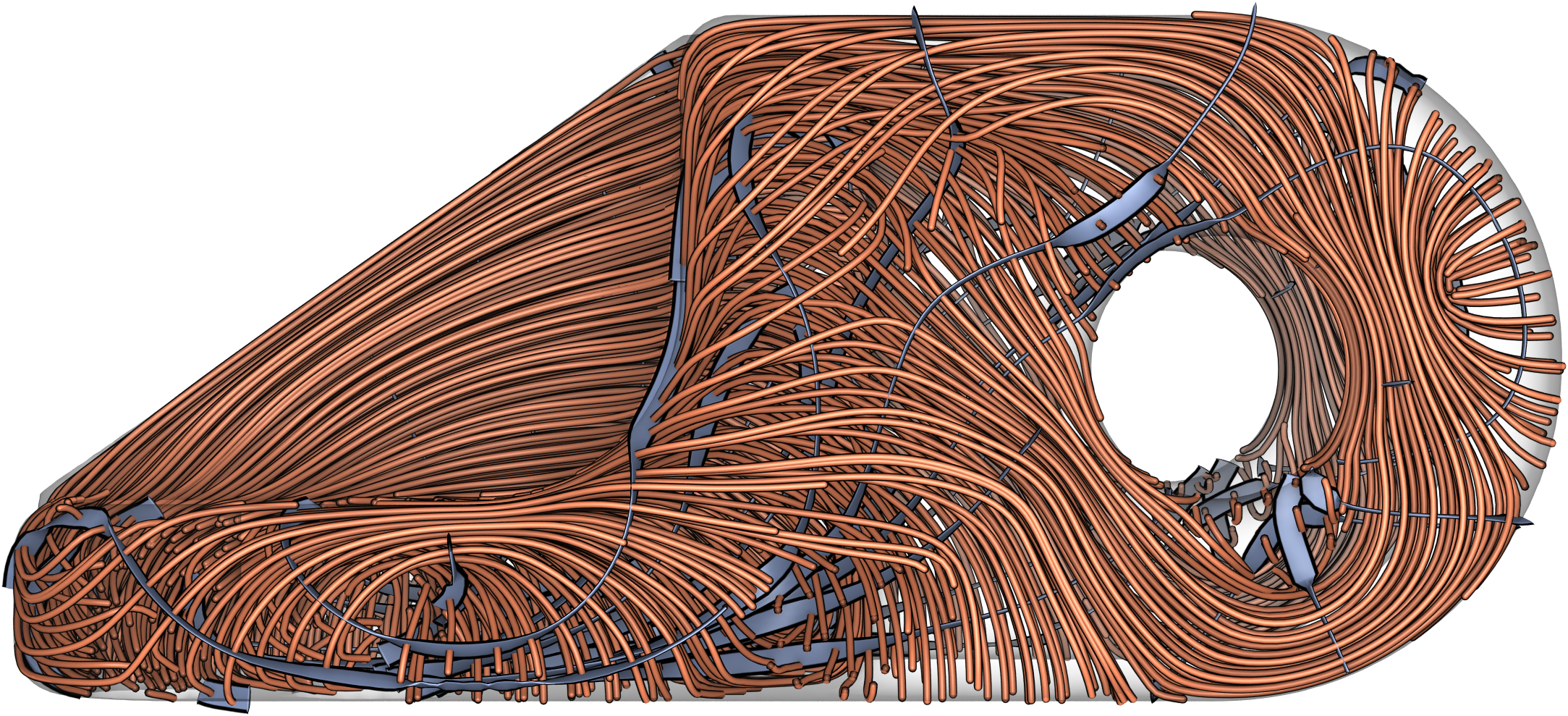} 
    \caption{Stress field in `Bearing'. Top: PSLs at the finest level (according to \autoref{tab:expCond}). Bottom: major / minor PSLs at the third ($L3$) / first ($L1$) level of detail. Ribbons are along the minor PSLs and twist according to medium principal stress direction.}
    \label{fig:bearing}
\end{figure}

\begin{figure}[ht]
    \centering
    \includegraphics[width=0.48\linewidth]{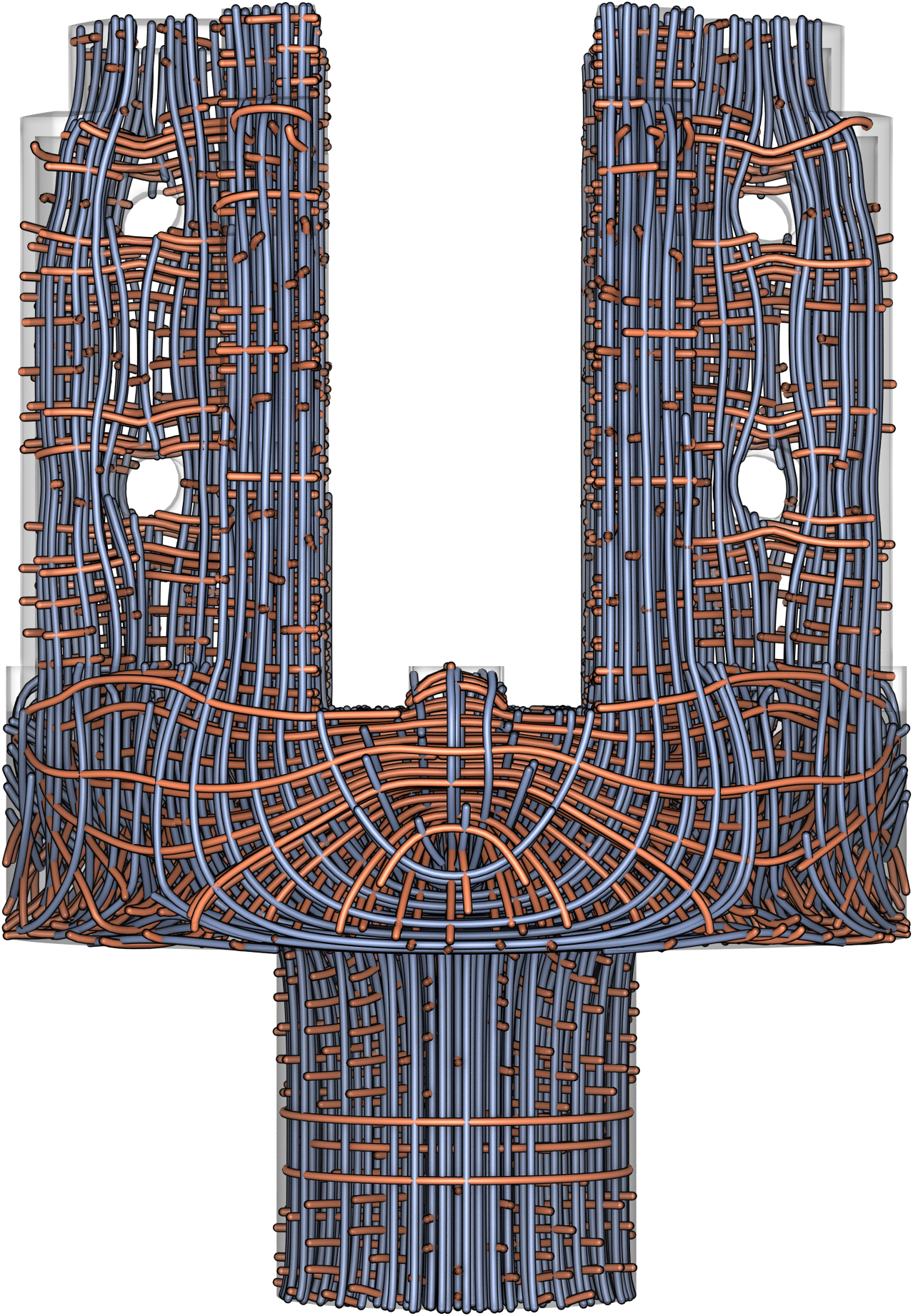}
    \includegraphics[width=0.48\linewidth]{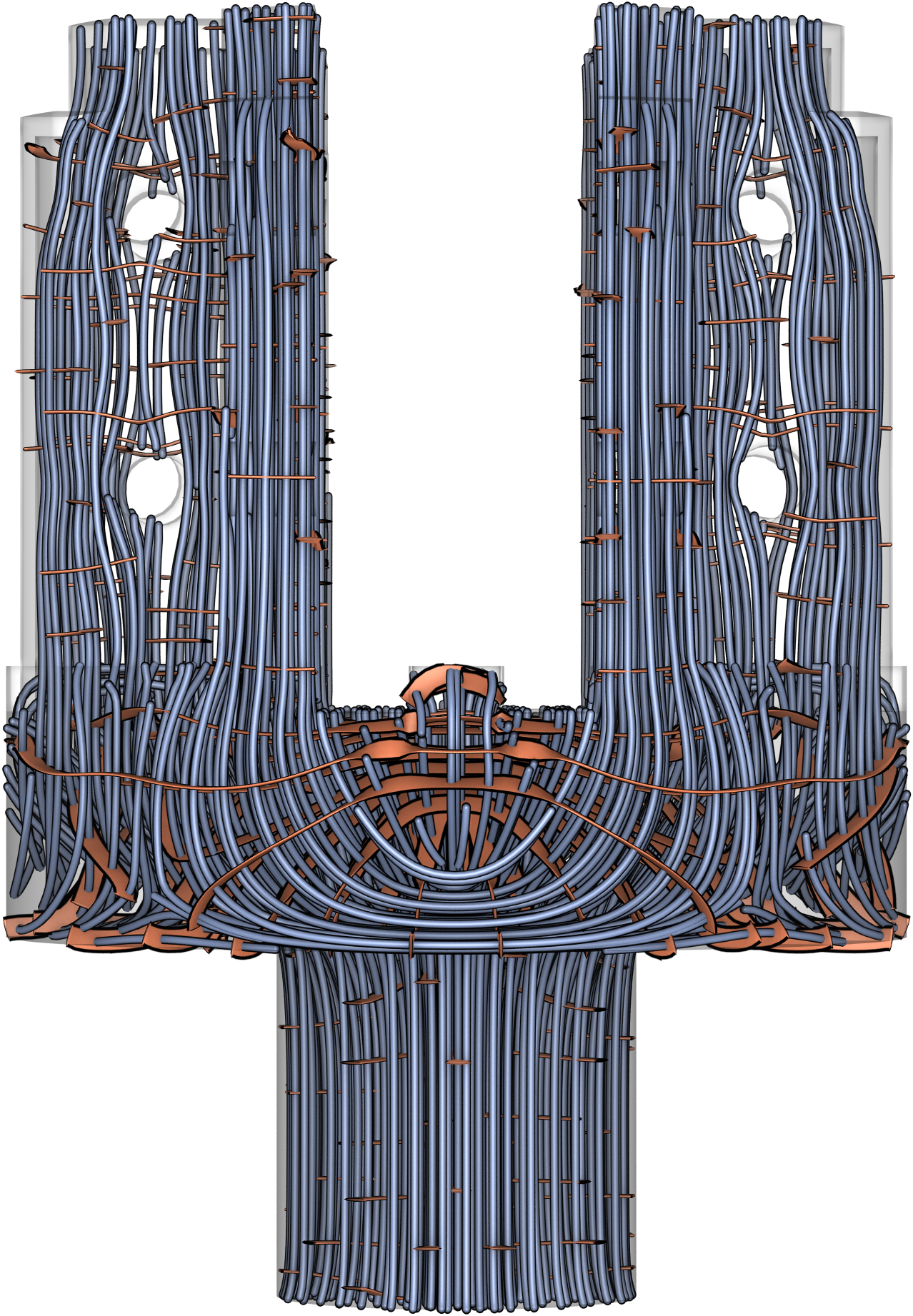}
    \caption{Stress fields in `Parts1'. Left: PSLs at the finest level. Right: major / minor PSLs at $L2$ / $L3$. Ribbons are along the major PSLs and twist according to medium principal stress direction.}
    \label{fig:parts1}
\end{figure}

3D-TSV works with Cartesian meshes and deformed hexahedral meshes, which are both frequently used in mechanical engineering applications. Here we use the stress fields due to external loads in the interior of `Bearing' and `Parts1', to demonstrate the capability of 3D-TSV. As shown in \autoref{fig:solidObjs}, especially in `Bearing' the element sizes change considerably over the 3D domain. The distribution of PSLs of `Bearing' is shown in \autoref{fig:bearing} (top), and the bottom image shows the combination of major at the third level of detail ($L3$) and minor at $L1$, where the minor PSLs are shown via ribbons. The full distribution of PSLs of `Parts1' can be seen in the \autoref{fig:parts1} (left), on the right the minor PSLs at $L3$ and major PSLs at $L2$ are shown simultaneously, where the major PSLs are rendered via ribbons.

\section{Conclusion and Future Work}
In this paper, we have introduced 3D-TSV, a tool for visualizing the principal stress directions in 3D solids under load. 3D-TSV makes use of a novel seeding strategy, to generate a space-filling and evenly spaced set of PSLs. By considering all three types of PSLs simultaneously in the construction process, the regularity of the resulting PSL structure is improved. By incorporating different merging thresholds for each PSL type into the construction process, a consistent multi-resolution hierarchy is formed, which can be utilized to show different PSL types with different resolutions simultaneously. 
Efficient rendering options for lines and ribbons on the GPU enable interactive analysis of large sets of PSLs. 

In the future, we intend to couple 3D-TSV with load simulation processes, so that dynamic changes of the stress field can be instantly monitored. Therefore, we will analyze whether the intrinsically iterative parts of the algorithm can be parallelized on modern multi-threading architectures. Furthermore, we are interested in using space-filling evenly spaced seeding to guide the material growth in topology optimization. Topology optimization seeks to distribute material in a way that makes the object resistant to external loads. To automatically generate support structures that follow the major stress directions and eventually can form a 3D grid-like structure, we aim at combining our seeding strategy with the automatic growth process underlying topology optimization.




\printcredits

\section*{Acknowledgment}
This work was supported in part by a grant from German Research Foundation (DFG) under grant number WE 2754/10-1. 
We acknowledge the help of Chunxiao Meng at Northwestern Polytechnical University and Yingjian Liu at The University of Texas at Dallas in adapting 3D-TSV to ANSYS and ABAQUS, respectively.

\bibliographystyle{plain} 

\bibliography{main}



\bio{figs/authors/junpeng}
Junpeng Wang is a PhD candidate in the Computer Graphics and Visualization Group at Technical University of Munich, Germany. He received his Bachelor and Master's degrees in Aerospace Science and Technology in 2015 and 2018, respectively, both from Northwestern Polytechnical University. Currently, his research is focused on tensor field visualization and numerical simulation for solid mechanics.
\endbio

\bio{figs/authors/christoph}
Christoph Neuhauser is a PhD candidate at the Computer Graphics and Visualization Group at the Technical University of Munich (TUM). He received his Bachelor's and Master's degrees in computer science from TUM in 2019 and 2020. Major interests in research comprise scientific visualization and real-time rendering.
\endbio

\bigskip

\bio{figs/authors/wu}
Jun Wu is an assistant professor at the Department of Sustainable Design Engineering, Delft University of Technology. Before this, he was a Marie Curie postdoc fellow at the Department of Mechanical Engineering, Technical University of Denmark. He obtained a PhD in Computer Science in 2015 from TUM, and a PhD in Mechanical Engineering in 2012 from Beihang University, Beijing. His research is focused on computational design and digital fabrication, with an emphasis on topology optimization.
\endbio

\bio{figs/authors/gao}
Xifeng Gao is currently a principal researcher with Tencent North America. He has more than 10 years of academic research experience. He is interested in solving geometric computing related problems in research areas, such as Computer Graphics, Digital Games, CAD/CAE, Multimedia Processing, Robotics, and Digital Fabrication.
\endbio

\bigskip
\newpage
\bio{figs/authors/ruediger}
R\"udiger Westermann studied computer science at the Technical University Darmstadt and received his Ph.D. in computer science from the University of Dortmund, both in Germany. In 2002, he was appointed the chair of Computer Graphics and Visualization at TUM. His research interests include scalable data visualization and simulation algorithms, GPU computing, real-time rendering of large data, and uncertainty visualization.
\endbio

\end{document}